\documentclass[aps,prd,twocolumn,showpacs,nofootinbib]{revtex4-1}

\usepackage{graphicx}
\usepackage{amsmath}
\usepackage{amssymb}
\usepackage{bm}
\usepackage{bbm}
\usepackage{color}
\usepackage{slashed}
\usepackage{multirow}
\usepackage[colorlinks=true,linkcolor=blue,citecolor=blue, urlcolor=blue]{hyperref}

\begin{document}

\title{Inclusive production of heavy quarkonium $\eta_Q$ via $Z$ boson decays within the framework of nonrelativistic QCD}

\author{Xu-Chang Zheng$^{a}$}
\email{zhengxc@cqu.edu.cn}
\author{Chao-Hsi Chang$^{b,c,d}$}
\email{zhangzx@itp.ac.cn}
\author{Xing-Gang Wu$^{a,e}$}
\email{wuxg@cqu.edu.cn}
\author{Xu-Dong Huang$^{a}$}
\email{hxud@cqu.edu.cn}
\author{Guang-Yu Wang$^{a}$}
\email{gywang@cqu.edu.cn}

\affiliation{$^a$ Department of Physics, Chongqing University, Chongqing 401331, P.R. China \\
$^b$ Key Laboratory of Theoretical Physics, Institute of Theoretical Physics, Chinese Academy of Sciences, Beijing 100190, P.R. China \\
$^c$ School of Physical Sciences, University of Chinese Academy of Sciences, Beijing 100049, P.R. China\\
$^d$ CCAST (World Laboratory), Beijing 100190, P.R. China \\
$^e$ Chongqing Key Laboratory for Strongly Coupled Physics, Chongqing 401331, P.R. China}

\begin{abstract}

In the paper, the inclusive production of heavy quarkonium $\eta_Q$ ($Q=b$ or $c$) via $Z$ boson decays within the framework of nonrelativistic QCD effective theory are studied. The contributions from the leading color-singlet and color-octet Fock states are considered. Total and differential decay widths for the inclusive decays $Z \to \eta_Q+X$ are presented. It is found that the decays $Z\to \eta_Q +X$ are dominated by the $^3S_1^{[8]}$ component, so the decays can be inversely adopted to determine the values of the long-distance matrix elements $\langle {\cal O}^{\eta_{c}}(^{3}S_{1}^{[8]})\rangle$ and $\langle {\cal O}^{\eta_{b}}(^{3}S_{1}^{[8]})\rangle$, respectively. Our numerical results show that at an $e^+e^-$ collider running at the $Z$ pole with a high luminosity around $10^{35}{\rm cm}^{-2}{\rm s}^{-1}$ (a super $Z$ factory), there are about $4.5\times 10^7$ $\eta_c$ meson events and $6.1\times 10^5$ $\eta_b$ meson events to be produced per operation year, and the inclusive decays may be used for clarifying some problems on the heavy quarkonium $\eta_Q$ and nonrelativistic QCD.

\end{abstract}


\maketitle

\section{Introduction}
\label{secIntro}

Heavy quarkonia have attracted a lot of interest since the discovery of the $J/\psi$ meson. An important reason is that they provide an ideal platform for studying the interplay between the perturbative and the nonperturbative effects in QCD. The nonrelativistic QCD (NRQCD) factorization formalism~\cite{nrqcd} provides a systematic framework to separate the short-distance and the long-distance effects in the heavy quarkonium production and decay processes. Under the NRQCD factorization, the heavy quarkonium production cross sections are expressed as the products of the short-distance coefficients (SDCs) and the long-distance matrix elements (LDMEs). The SDCs describe the production of heavy quark-antiquark pairs with proper quantum numbers, which can be calculated perturbatively. The LDMEs describe the hadronization of a produced heavy quark pair into quarkonium, which are nonperturbative in nature but can be extracted from a global fit of experimental measurements or estimated by using the QCD inspired potential models etc.

Up to now, the NRQCD factorization formalism has achieved great successes in explaining the data at the high-energy colliders~\cite{ybook1, ybook2}. However, there are still some challenges. For instance, the global fits of the $J/\psi$ color-octet (CO) LDMEs from various groups are not so consistent with each other, cf. Refs.\cite{Butenschoen:2011yh, Chao:2012iv, Gong:2012ug, Brambilla:2014jmp}. Thus, it is interesting to study more quarkonium processes relating the NRQCD factorization formalism.

Most studies of the quarkonia focus on the $J/\psi$ and $\Upsilon$ mesons due to their high detection efficiency. For instance, the $J/\psi$ events can be reconstructed via the decays $J/\psi \to l^+ l^-(l=e,\mu)$ with high efficiency, whose total branching ratio is $\sim 12\%$ \cite{Zyla:2020zbs}. Contrary to the $J/\psi$ meson, there are less studies of the $\eta_c$ meson production. Conventionally, the decay channel used to reconstruct the $\eta_c$ events is $\eta_c \to \gamma \gamma$, and the branching ratio of this decay channel is $\sim 1.6\times 10^{-4}$~\cite{Zyla:2020zbs}. Moreover, it is very difficult to record the two photons from the background in a hadron collision environment. Namely the experimental detection of the $\eta_c$ meson is poor. A novel proposal to reconstruct the $\eta_c$ events through the decay channel $\eta_c \to p\bar{p}$ has been suggested in Ref.\cite{Barsuk:2012ic}, whose branching ratio is $\sim 1.5\times 10^{-3}$~\footnote{In fact, the decay channels $\eta_c \to \Lambda\bar{\Lambda}$ and $\eta_c \to \Sigma\bar{\Sigma}$, whose branching ratios are $\sim 1.07\times10^{-3}$ and $\sim 2.1\times10^{-3}$, respectively \cite{Zyla:2020zbs}, may also be used to identify the $\eta_c$ meson so as to increase the detection efficiency of $\eta_c$. It is not very difficult to detect the strange baryon pairs produced from the $\eta_c$ decay with vertex detectors because they carry high momentum from the $\eta_c$ and make tracks.}. This proposal opened a new way to study the $\eta_c$ meson at the high-energy colliders, and it has been adopted to observe the $\eta_c$ meson by the LHCb Collaboration~\cite{Aaij:2014bga, Aaij:2019gsn}. Recent theoretical studies of the $\eta_c$ production at the LHC can be found in Refs.\cite{Butenschoen:2014dra, Han:2014jya, Zhang:2014ybe, Goncalves:2018yxc, Feng:2019zmn, Baranov:2019joi, Babiarz:2019mag, Tichouk:2020zhh, Tichouk:2020dut}.

The $\eta_b$ meson has the same quantum numbers as those of the $\eta_c$ meson, but has different constituent quark mass. Since the heavier bottom quark mass, the $\eta_b$ meson is a better object for applying NRQCD. Thus it is interesting to study the $\eta_b$ and $\eta_c$ production applying the NRQCD factorization at the same time, although the observations on the $\eta_b$ are scarce. Up to now, the $\eta_b$ has been observed only through the feed-down contributions, i.e., from the decays of excited bottonium states. Therefore, the studies of the $\eta_b$ production from various processes are requested.

At the LHC or an $e^+ e^-$ collider running around the $Z$ pole and with an accessible high-luminosity (a super $Z$ factory), the production of the heavy quarkonium $\eta_Q$ through $Z$ boson decays can provide abundant information. The inclusive $Z$ boson production cross section at the LHC with the collision energy $13\,{\rm TeV}$ is $\sim 56$ nb~\cite{Law:2016hqt}. With the luminosity of $10^{34}{\rm cm}^{-2}{\rm s}^{-1}$, there are $\sim 5.6\times 10^9$ $Z$ bosons to be produced per operation year at the LHC. A Chinese group has proposed to build a super $Z$ factory~\cite{zfactory}, and its luminosity of the super $Z$ factory could reach to $10^{34- 36}{\rm cm}^{-2}{\rm s}^{-1}$, which is higher than that of the LEP-I by three to five orders. The $Z$ boson production cross section is $\sim 30$ nb, and there are about $3\times 10^{9- 11}$ $Z$ bosons to be produced per operation year at the super $Z$ factory. Therefore, it is interesting to study the production of heavy quarkonia through $Z$ boson decays.

The production of heavy quarkonia ($J/\psi$ and $\Upsilon$ etc) through $Z$ decays has been extensively studied at the leading order in $\alpha_s$ and $v_Q$ \cite{Guberina:1980dc, Keung:1980ev, Abraham:1989ri, Barger:1989cq, Hagiwara:1991mt, Braaten:1993mp, Fleming:1993fq, Liao:2015vqa}, the typical velocity of the heavy quark in quarkonia. For the $J/\psi$ and $\Upsilon$ production through $Z$ boson decays, the CO contributions have been estimated in Refs.\cite{Ernstrom:1996aa, Schuler:1997is, Cheung:1995ka, Cho:1995vv}, the next-to-leading-order (NLO) QCD corrections have been calculated in Ref.\cite{JXWang}, and the leading and next-to-leading logarithms of $m_{_Z}/m_Q$ have been resummed through the fragmentation approach in Ref.\cite{jpsiFFNLO}. In the present paper, we devote ourselves to studying the inclusive production of $\eta_Q$ with $Q=c$ or $b$ through the $Z$ boson decays.

According to NRQCD, for the $\eta_Q$ production, the leading color-singlet (CS) and color-octet (CO) Fock states are $^1S_0^{[1]}$ at $v_Q^3$ order and $^1S_0^{[8]}$, $^3S_1^{[8]}$, and $^1P_1^{[8]}$ at $v_Q^7$ order. Although the CO contributions are suppressed by $v_Q^4$ order compared to the CS contribution in the long-distance part, the CO contributions may be enhanced in the short-distance part. Therefore, besides the CS state $^1S_0^{[1]}$, we also consider the CO states $^1S_0^{[8]}$, $^3S_1^{[8]}$, and $^1P_1^{[8]}$.

The remaining parts of the paper are organized as follows. In Sec.\ref{caltech}, we briefly present the useful formulas for calculating the $Z\to\eta_Q$ inclusive decays. In Sec.\ref{secNumer}, numerical results are presented. Section \ref{secSum} is reserved for discussion and conclusion.

\section{Calculation technology}
\label{caltech}

Under the NRQCD factorization formalism, the decay width for the inclusive process $Z \to \eta_Q+X$ can be written as
\begin{eqnarray}
 d\Gamma_{Z\to \eta_Q+X}=\sum_n d\tilde{\Gamma}_{Z\to (Q\bar{Q})[n]+X}\langle {\cal O}^{\eta_Q}(n)\rangle,\label{nrqcdfact}
\end{eqnarray}
where $d\tilde{\Gamma}$ are the perturbatively calculable SDCs and $\langle {\cal O}^{\eta_Q}(n)\rangle$ are the nonperturbative LDMEs. The sum extends over the intermediate states $^{2S+1}L_{J}^{[1,8]}$. Up to relative $v_Q^4$ order, the LDMEs $\langle {\cal O}^{\eta_Q}(^1S_0^{[1]})\rangle$, $\langle {\cal O}^{\eta_Q}(^1S_0^{[8]})\rangle$, $\langle {\cal O}^{\eta_Q}(^3S_1^{[8]})\rangle$, and $\langle {\cal O}^{\eta_Q}(^1P_1^{[8]})\rangle$ are involved.

To calculate the decay width for $Z \to \eta_Q+X$, we first calculate the decay widths for a free on shell $(Q\bar{Q})$ pair with the quantum numbers $^{2S+1}L_{J}^{[1,8]}$, i.e., $d\Gamma_{Z\to (Q\bar{Q})[^{2S+1}L_{J}^{[1,8]}]+X}$. Then the contributions of different channels to the decay width of the $Z \to \eta_Q+X$ are obtained from $d\Gamma_{Z\to (Q\bar{Q})[^{2S+1}L_{J}^{[1,8]}]+X}$ through replacing the matrix element $\langle {\cal O}^{(Q\bar{Q})[^{2S+1}L_{J}^{[1,8]}]}(^{2S+1}L_{J}^{[1,8]})\rangle$ by $\langle {\cal O}^{\eta_Q}(^{2S+1}L_{J}^{[1,8]})\rangle$.

In the paper, we consider the contributions from the processes up to $\alpha \alpha_s^2$ order. The involved decay channels are
\begin{eqnarray}
&& Z \to \eta_Q(^1S_0^{[8]}, ^3S_1^{[8]}, ^1P_1^{[8]})+g,   \\
&& Z \to \eta_Q(^1S_0^{[1]}, ^1S_0^{[8]}, ^3S_1^{[8]}, ^1P_1^{[8]})+gg,   \\
&& Z \to \eta_Q(^1S_0^{[8]}, ^3S_1^{[8]}, ^1P_1^{[8]})+q\bar{q}, \\
&& Z \to \eta_Q(^1S_0^{[1]}, ^1S_0^{[8]}, ^3S_1^{[8]}, ^1P_1^{[8]})+Q\bar{Q},  \\
&& Z \to \eta_Q(^1S_0^{[8]}, ^3S_1^{[8]}, ^1P_1^{[8]})+Q'\bar{Q'} \; (Q'\neq Q).
\end{eqnarray}
The decay channels $Z \to \eta_Q(^1S_0^{[8]}, ^3S_1^{[8]}, ^1P_1^{[8]})+gg$ and $Z \to \eta_Q(^1S_0^{[8]}, ^3S_1^{[8]}, ^1P_1^{[8]})+q\bar{q}$ are the real corrections to the decay channels $Z \to \eta_Q(^1S_0^{[8]}, ^3S_1^{[8]}, ^1P_1^{[8]})+g$, and should be considered together with the virtual corrections to the decay channels $Z \to \eta_Q(^1S_0^{[8]}, ^3S_1^{[8]}, ^1P_1^{[8]})+g$ so as to obtain finite predictions.

The decay width for the $(Q\bar{Q})[^{2S+1}L_{J}^{[1,8]}]$ pair can be written as
\begin{eqnarray}
d\Gamma_{Z \to(Q\bar{Q})[^{2S+1}L_J^{[1,8]}]+X}=\frac{1}{3}\frac{1}{2m_{_Z}}\sum \vert {\cal M} \vert^2 d\Phi_N,
\end{eqnarray}
where $N$ indicates that there are $N$ particles in the final state, $\sum$ indicates the sum over the spin and color states of initial and final particles, and $1/3$ comes from the polarization average of the initial $Z$ boson. $d\Phi_N$ is $N$-body differential phase space
\begin{eqnarray}
d\Phi_N=(2\pi)^4 \delta^4 \left(p_0-\sum_{f=1}^N p_f\right)\prod_{f=1}^N \frac{d^{3} \textbf{p}_f}{(2\pi)^{3} 2E_f}, \label{dphi3}
\end{eqnarray}
and ${\cal M}$ denotes the amplitude for the $(Q\bar{Q})[^{2S+1}L_{J}^{[1,8]}]$ pair. In the following, we sketch the formulas used in the calculation of these decay channels, successively.

\subsection{$Z \to \eta_Q(^1S_0^{[8]}, ^3S_1^{[8]}, ^1P_1^{[8]}) +g$ and their NLO QCD corrections}

\subsubsection{Leading order contributions}

\begin{figure}[htbp]
\includegraphics[width=0.45\textwidth]{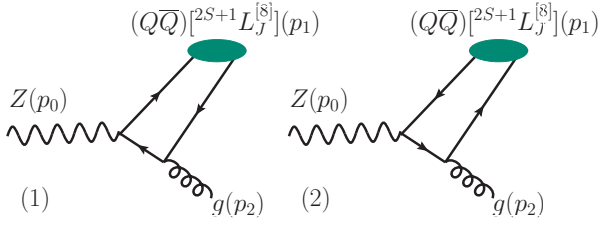}
\caption{Feynman diagrams for $Z \to (Q\bar{Q})[^{2S+1}L_J^{[8]}]+g$.
 } \label{feyn-g}
\end{figure}

At leading order (LO) in $\alpha_s$, there are two Feynman diagrams for the decay channel $Z \to (Q\bar{Q})[^{2S+1}L_J^{[8]}]+g$, which are shown in Fig.\ref{feyn-g}. The amplitude (${\cal M}={\cal M}_1+{\cal M}_2$) for the decay channel can be written down according to the two Feynman diagrams. For the $^1S_0^{[8]}$ ($^3S_1^{[8]}$) case, we have
\begin{eqnarray}
i {\cal M}_1=&&-\frac{i g}{2{\rm cos}\, \theta_W}{\rm tr}\Big[\Pi_{1(3)} \Lambda^a_8\slashed{\epsilon}(p_0) (V_Q-A_Q\gamma_5)\nonumber  \\
&& \cdot  \frac{i}{-\slashed{p}_{12}-\slashed{p}_2-m_Q+i \epsilon}  (ig_s \slashed{\epsilon}^*(p_2)T^b)\Big] \Big\vert_{q=0}, \\
i {\cal M}_2=&&-\frac{i g}{2{\rm cos}\, \theta_W}{\rm tr}\Big[\Pi_{1(3)} \Lambda^a_8 (ig_s \slashed{\epsilon}^*(p_2)T^b)\nonumber  \\
&& \cdot  \frac{i}{\slashed{p}_{11}+\slashed{p}_2-m_Q+i \epsilon} \slashed{\epsilon}(p_0) (V_Q-A_Q\gamma_5) \Big] \Big\vert_{q=0},
\end{eqnarray}
where $V_Q$ and $A_Q$ are vector and axial electroweak couplings. More explicitly, $V_Q=T_{3Q}-2 e_Q \,{\rm sin}^2\theta_W$ and $A_Q=T_{3Q}$, where $T_{3Q}$ and $e_Q$ are weak isospin and the charge of fermion $Q$ in units of positron charge, respectively. $p_{11}=p_1/2+q$ and $p_{12}=p_1/2-q$ are the momenta of the $Q$ and $\bar{Q}$ in the $(Q\bar{Q})[^{2S+1}L_J^{[8]}]$ pair, and $\Pi_{1(3)}$ is the spin-singlet (spin-triplet) projector, i.e.,
\begin{eqnarray}
&&\Pi_1=\frac{1}{(2\,m_Q)^{3/2}}(\slashed{p}_{12}-m_Q)\gamma_5(\slashed{p}_{11}+m_Q),\\
&&\Pi_3=\frac{1}{(2\,m_Q)^{3/2}}(\slashed{p}_{12}-m_Q)\slashed{\epsilon}^*(p_1)(\slashed{p}_{11}+m_Q),
\end{eqnarray}
and $\Lambda^a_8=\sqrt{2}T^a$ is the CO projector. For the $^1P_1^{[8]}$ case, we have
\begin{eqnarray}
i {\cal M}_1=&&-\frac{i g\,\epsilon^*_{\alpha}(p_1)}{2{\rm cos}\, \theta_W}\frac{d}{dq_{\alpha}}{\rm tr}\Big[\Pi_{1} \Lambda^a_8\slashed{\epsilon}(p_0) (V_Q-A_Q\gamma_5)\nonumber  \\
&& \cdot  \frac{i}{-\slashed{p}_{12}-\slashed{p}_2-m_Q+i \epsilon}  (ig_s \slashed{\epsilon}^*(p_2)T^b)\Big] \Big\vert_{q=0}, \\
i {\cal M}_2=&&-\frac{i g\,\epsilon^*_{\alpha}(p_1)}{2{\rm cos}\, \theta_W}\frac{d}{dq_{\alpha}}{\rm tr}\Big[\Pi_{1} \Lambda^a_8 (ig_s \slashed{\epsilon}^*(p_2)T^b)\nonumber  \\
&& \cdot  \frac{i}{\slashed{p}_{11}+\slashed{p}_2-m_Q+i \epsilon} \slashed{\epsilon}(p_0) (V_Q-A_Q\gamma_5) \Big] \Big\vert_{q=0}.
\end{eqnarray}
Squaring the amplitude ${\cal M}$ and integrating the squared amplitude over the two-body phase space, we obtain the LO contribution for the decay channel $Z \to (Q\bar{Q})[^{2S+1}L_J^{[8]}]+g$.

\subsubsection{Virtual corrections}

\begin{figure}[htbp]
\includegraphics[width=0.45\textwidth]{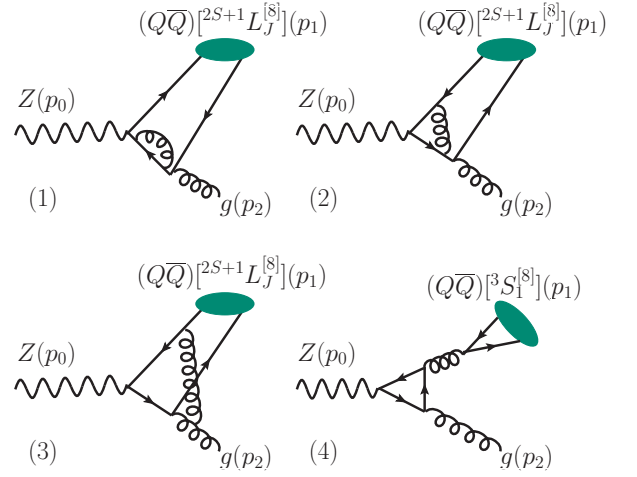}
\caption{Four sample Feynman diagrams for the virtual corrections to $Z \to (Q\bar{Q})[^{2S+1}L_J^{[8]}]+g$.
 } \label{feyn-g-vir}
\end{figure}

The NLO virtual corrections come from the interference of the one-loop diagrams and the LO diagrams. Four sample one-loop Feynman diagrams are shown in Fig.\ref{feyn-g-vir}. The fourth sample Feynman diagram is specific to the $^3S_1^{[8]}$ channel. The amplitude is too lengthy to be listed here.

There are UV divergences in the self-energy and vertex diagrams and IR divergences in the vertex and box diagrams. We adopt dimensional regularization with $d=4-2\epsilon$ to regularize these divergences. Then the divergences appear as pole terms in $\epsilon$. The $\gamma_5$ matrix should be noted in dimensional regularization, and we adopt the reading point prescription \cite{Korner:1991sx} to deal with it.

The UV divergences should be removed through renormalization. In the calculation, the renormalization scheme is adopted as follows: the renormalization of the quark field, the quark mass and the gluon field is carried out in the on-mass-shell (OS) scheme, while the renormalization of the strong coupling constant is carried out in the modified minimal subtraction ($\overline{\rm MS}$) scheme. The renormalization constants are
\begin{eqnarray}
 \delta Z^{\rm OS}_{2,Q}&=&-C_F \frac{\alpha_s}{4\pi}\left[\frac{1}{\epsilon_{UV}}+ \frac{2}{\epsilon_{IR}}-3~\gamma_E+3~ {\rm ln}\frac{4\pi \mu_R^2}{m_Q^2}+4\right], \nonumber\\
\delta Z^{\rm OS}_{m,Q}&=&-3~C_F \frac{\alpha_s}{4\pi}\left[\frac{1}{\epsilon_{UV}}- \gamma_E+
 {\rm ln}\frac{4\pi \mu_R^2}{m_Q^2}+\frac{4}{3}\right], \nonumber\\
 \delta Z^{\rm OS}_3&=&\frac{\alpha_s}{4\pi}\left[(\beta'_0-2C_A) \left(\frac{1}{\epsilon_{UV}}-\frac{1}{\epsilon_{IR}}\right) \right. \nonumber\\
 && \left.-\frac{4}{3}T_F \sum_Q \left(\frac{1}{\epsilon_{UV}}-\gamma_E + {\rm ln}\frac{4\pi \mu_R^2}{m_Q^2}\right)\right], \nonumber\\
 \delta Z^{\overline{\rm MS}}_g&=&- \frac{\beta_0}{2}\frac{\alpha_s}{4\pi}\left[\frac{1}{\epsilon_{UV}}- \gamma_E+ {\rm ln}~(4\pi) \right], \nonumber
\end{eqnarray}
where $\gamma_E$ is the Euler constant, $\mu_R$ is the renormalization scale, $\beta_0=11-2 n_f/3$ is the one-loop coefficient of the QCD $\beta$ function, and $n_f$ is the flavor number of active quarks. $\beta'_0=11-2 n_{lf}/3$ and $n_{lf}=3$ is the number of light-quark flavors. For $SU(3)$ group, $C_F=4/3$, $T_F=1/2$, and $C_A=3$.

In the calculation, the threshold expansion method \cite{region} is employed to extract the SDCs, i.e., we expand the relative momentum ($q$) of the $(Q\bar{Q})[^{2S+1}L_J^{[8]}]$ pair before performing the loop integration. Then the Coulomb divergences, which are IR power divergences and vanish in dimensional regularization, do not appear in our calculation.

\subsubsection{Real corrections}

\begin{figure}[htbp]
\includegraphics[width=0.45\textwidth]{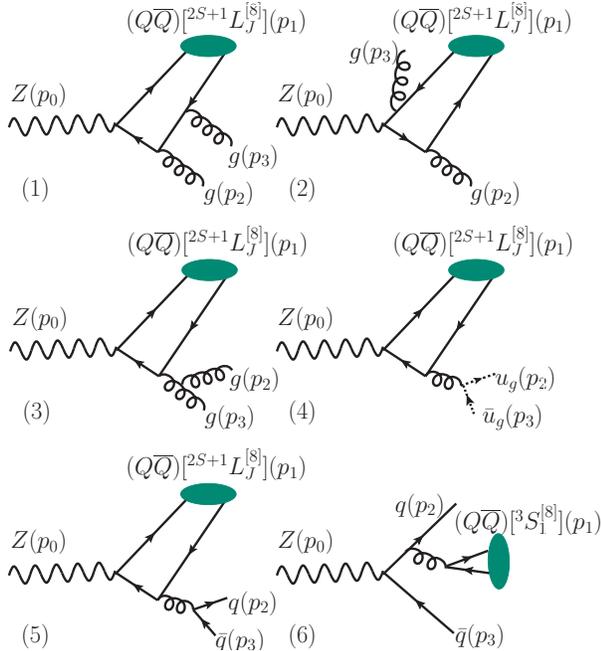}
\caption{Six sample Feynman diagrams for the real corrections to $Z \to (Q\bar{Q})[^{2S+1}L_J^{[8]}]+g$.
 } \label{feyn-g-real}
\end{figure}

The real corrections to the decay channel $Z \to (Q\bar{Q})[^{2S+1}L_J^{[8]}]+g$ come from the processes $Z \to (Q\bar{Q})[^{2S+1}L_J^{[8]}]+gg$ and $Z \to (Q\bar{Q})[^{2S+1}L_J^{[8]}]+q\bar{q}$, where $q=u,d,s$. In the calculation, we use $\sum_{i} \epsilon^{\mu*}_i\epsilon^{\nu}_i\to -g_{\mu\nu}$ to sum the polarizations of the final-state gluons. The unphysical polarization contributions are subtracted through the process involving ghost-pair production, i.e., $Z \to (Q\bar{Q})[^{2S+1}L_J^{[8]}]+u_g \bar{u}_g$. Six sample Feynman diagrams for the real corrections are shown in Fig.\ref{feyn-g-real}. The sixth diagram is specific to the $^3S_1^{[8]}$ channel.

There are IR divergences in the real corrections. These IR divergences should be regularized by dimensional regularization as those in the virtual corrections. In order to simplify the calculation of the real corrections under dimensional regularization, we adopt the two-cutoff phase-space slicing method \cite{twocutoff} to isolate the divergent terms. Under this method, the differential decay width for the real corrections can be decomposed into three parts,
\begin{eqnarray}
d \Gamma_{\rm Real}=d \Gamma_{\rm S}+d \Gamma_{\rm HC}+d \Gamma_{\rm H\bar{C}},
\end{eqnarray}
where $d \Gamma_{\rm S}$ denotes the contribution from the phase space region with $E_{2}\leq m_{_Z} \delta_s/2$ or $E_{3}\leq m_{_Z} \delta_s/2$, $d \Gamma_{\rm HC}$ denotes the contribution from the phase space region with $E_{2}> m_{_Z} \delta_s/2$, $E_{3}> m_{_Z} \delta_s/2$ and $(p_2+p_3)^2 \leq m_{_Z}^2 \delta_c$, and $\Gamma_{\rm H\bar{C}}$ denotes the contribution from the phase space region with $E_{2}> m_{_Z} \delta_s/2$, $E_{3}> m_{_Z} \delta_s/2$ and $(p_2+p_3)^2 > m_{_Z}^2 \delta_c$. In the calculation, the two cutoff parameters should be taken as $\delta_c\ll \delta_s\ll 1$. Applying the eikonal and collinear approximations to the soft and hard-collinear parts respectively, $\Gamma_{\rm S}$ and $\Gamma_{\rm HC}$ can be calculated analytically in $d$ space-time dimensions. Due to the constraints $E_{2}> m_{_Z} \delta_s/2$, $E_{3}> m_{_Z} \delta_s/2$ and $(p_2+p_3)^2 > m_{_Z}^2 \delta_c$, $\Gamma_{\rm H\bar{C}}$ is finite and can be calculated in four space-time dimensions safely.

After summing the virtual and real corrections, the IR divergences are canceled in the $^{1}S_0^{[8]}$ and $^{3}S_1^{[8]}$ cases. However, there are IR divergences remaining in the $^{1}P_1^{[8]}$ case after summing the virtual and real corrections. The finite SDC for the $^{1}P_1^{[8]}$ channel can extracted through matching.

Applying the NRQCD factorization to the production of an on shell $(Q\bar{Q})[^{1}P_1^{[8]}]$ pair, the decay width can be written as
\begin{eqnarray}
d \Gamma_{^{1}P_1^{[8]}}=&& d \tilde{\Gamma}_{^{1}P_1^{[8]}}\langle {\cal O}^{(Q\bar{Q})[^{1}P_{1}^{[8]}]}(^{1}P_{1}^{[8]})\rangle\nonumber \\
&& +d \tilde{\Gamma}_{^{1}S_0^{[1]}}\langle {\cal O}^{(Q\bar{Q})[^{1}P_{1}^{[8]}]}(^{1}S_{0}^{[1]})\rangle \nonumber \\
&& +d \tilde{\Gamma}_{^{1}S_0^{[8]}}\langle {\cal O}^{(Q\bar{Q})[^{1}P_{1}^{[8]}]}(^{1}S_{0}^{[8]})\rangle,
\end{eqnarray}
where $d \Gamma_{^{1}P_1^{[8]}}$ denotes the decay width for an on shell $(Q\bar{Q})$ pair with quantum numbers $^{1}P_1^{[8]}$, $\langle {\cal O}^{(Q\bar{Q})[^{1}P_{1}^{[8]}]}(^{2S+1}L_{J}^{[8]})\rangle$ denotes the LDMEs for the quark pair. The LDME $\langle {\cal O}^{(Q\bar{Q})[^{1}P_{1}^{[8]}]}(^{1}P_{1}^{[8]})\rangle$ starts at $\alpha_s^0$ order, while $\langle {\cal O}^{(Q\bar{Q})[^{1}P_{1}^{[8]}]}(^{1}S_{0}^{[1]})\rangle$ and $\langle {\cal O}^{(Q\bar{Q})[^{1}P_{1}^{[8]}]}(^{1}S_{0}^{[8]})\rangle$ start at $\alpha_s$ order. Up to $\alpha_s^2$ order, the second term vanishes in this decay channel. Then, we have
\begin{eqnarray}
 d \tilde{\Gamma}_{^{1}P_1^{[8]}}=&& \frac{ d \Gamma_{^{1}P_1^{[8]}}}{\langle {\cal O}^{(Q\bar{Q})[^{1}P_{1}^{[8]}]}(^{1}P_{1}^{[8]})\rangle} \nonumber \\
&&-\frac{d \tilde{\Gamma}_{^{1}S_0^{[8]}}\langle {\cal O}^{(Q\bar{Q})[^{1}P_{1}^{[8]}]}(^{1}S_{0}^{[8]})\rangle}{\langle {\cal O}^{(Q\bar{Q})[^{1}P_{1}^{[8]}]}(^{1}P_{1}^{[8]})}.
\end{eqnarray}
Under the $\overline{\rm MS}$ factorization scheme,
\begin{eqnarray}
&&\langle {\cal O}^{(Q\bar{Q})[^{1}P_{1}^{[8]}]}(^{1}S_{0}^{[8]})\rangle\nonumber \\
&&=-\frac{\alpha_s B_F C_{\epsilon}}{3\pi m_Q^2} \langle {\cal O}^{(Q\bar{Q})[^{1}P_{1}^{[8]}]}(^{1}P_{1}^{[8]})\rangle,
\end{eqnarray}
where $B_F=(N_c^2-4)/N_c$ and $C_{\epsilon}=1/\epsilon_{IR}-\gamma_E+{\rm ln}(4\pi)+{\rm ln}(\mu_R^2/\mu_\Lambda^2)$. Then the finite SDC $d \tilde{\Gamma}_{^{1}P_1^{[8]}}$ is obtained.

\subsection{$Z \to \eta_Q(^1S_0^{[1]},^1S_0^{[8]},^3S_1^{[8]},^1P_1^{[8]}) +Q\bar{Q}$}

\begin{figure}[htbp]
\includegraphics[width=0.45\textwidth]{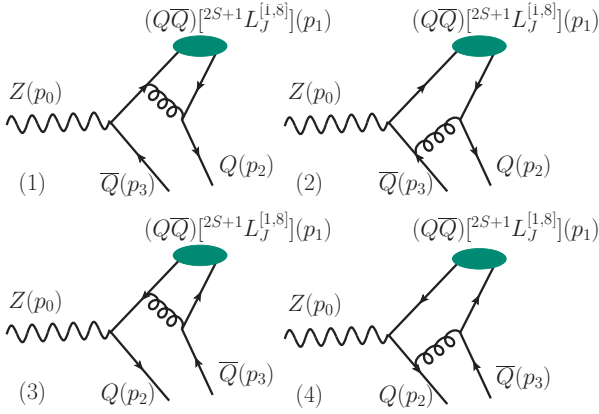}
\caption{Four of the Feynman diagrams for $Z \to (Q\bar{Q})[^{2S+1}L_J^{[1,8]}]+Q\bar{Q}$.
 } \label{feyn-QQ-1}
\end{figure}

\begin{figure}[htbp]
\includegraphics[width=0.45\textwidth]{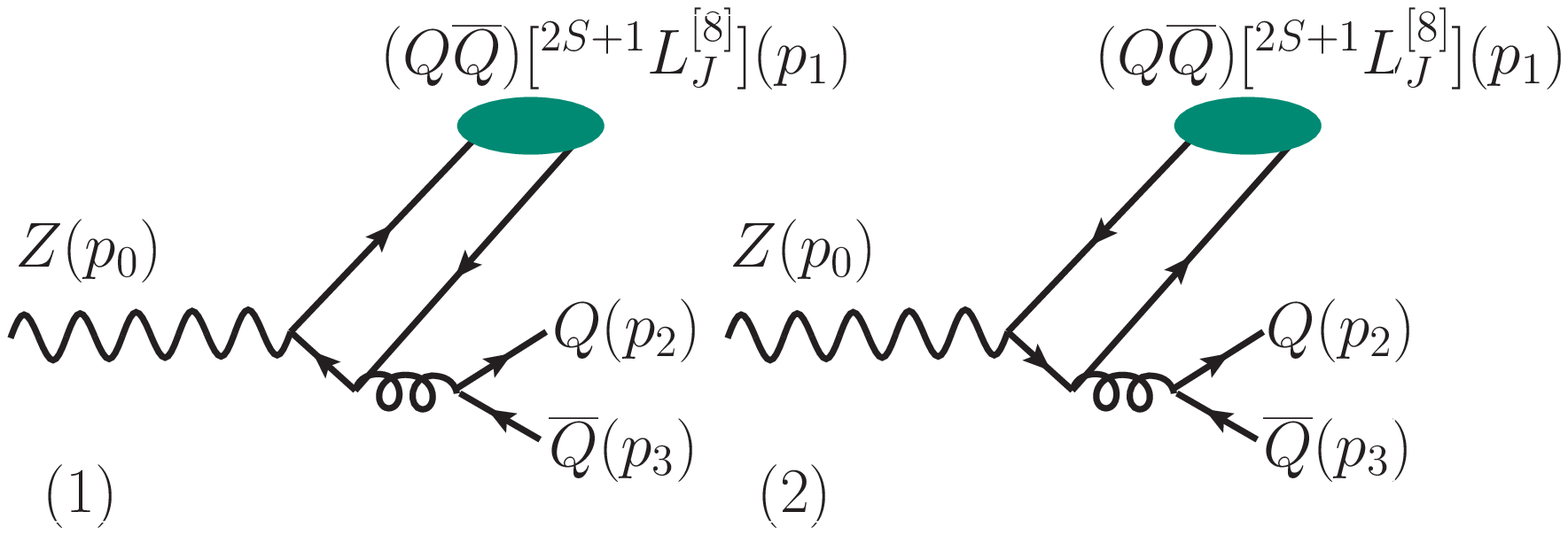}
\caption{Two of the Feynman diagrams for $Z \to (Q\bar{Q})[^{2S+1}L_J^{[8]}]+Q\bar{Q}$.
 } \label{feyn-QQ-2}
\end{figure}

For the CS decay channel $Z \to (Q\bar{Q})[^{1}S_0^{[1]}]+Q\bar{Q}$, there are four Feynman diagrams which are shown in Fig.\ref{feyn-QQ-1}. The amplitude (${\cal M}=\sum_{i=1}^{4}{\cal M}_i$) can be written down according to the four Feynman diagrams,
\begin{eqnarray}
i{\cal M}_1=&&-\frac{i g}{2{\rm cos}\, \theta_W}\frac{-i}{(p_{12}+p_2)^2+i\epsilon}\bar{u}(p_2)(ig_s \gamma^{\mu}T^b) \nonumber \\
&& \cdot \Pi_1 \Lambda_1 (ig_s \gamma_{\mu}T^b) \frac{i}{\slashed{p}_1+\slashed{p}_2-m_Q+i \epsilon} \slashed{\epsilon}(p_0)  \nonumber \\
&& \cdot (V_Q-A_Q\gamma_5)v(p_3)\big{\vert}_{q=0}, \label{eq.QQ1}\\
i{\cal M}_2=&&-\frac{i g }{2{\rm cos}\, \theta_W}\frac{-i}{(p_{12}+p_2)^2+i\epsilon}\bar{u}(p_2)(ig_s \gamma^{\mu}T^b) \Pi_1 \nonumber \\
&&  \cdot \Lambda_1 \slashed{\epsilon}(p_0)(V_Q-A_Q\gamma_5)\frac{i}{-\slashed{p}_0+\slashed{p}_{11}-m_Q+i \epsilon} \nonumber \\
&&.(ig_s \gamma_{\mu}T^b)v(p_3)\big{\vert}_{q=0},\label{eq.QQ2}
\end{eqnarray}
\begin{eqnarray}
i{\cal M}_3=&&-\frac{i g }{2{\rm cos}\, \theta_W}\frac{-i}{(p_{11}+p_3)^2+i\epsilon}\bar{u}(p_2)\slashed{\epsilon}(p_0) \nonumber \\
&&  \cdot (V_Q-A_Q\gamma_5) \frac{i}{-\slashed{p}_1-\slashed{p}_3-m_Q+i \epsilon} (ig_s \gamma_{\mu}T^b) \nonumber \\
&&.\Pi_1 \Lambda_1 (ig_s \gamma^{\mu}T^b) v(p_3)\big{\vert}_{q=0},\label{eq.QQ3}\\
i{\cal M}_4=&&-\frac{i g }{2{\rm cos}\, \theta_W}\frac{-i}{(p_{11}+p_3)^2+i\epsilon}\bar{u}(p_2)(ig_s \gamma_{\mu}T^b)  \nonumber \\
&& \cdot \frac{i}{\slashed{p}_0-\slashed{p}_{12}-m_Q+i \epsilon} \slashed{\epsilon}(p_0)(V_Q-A_Q\gamma_5)  \nonumber \\
&&.\Pi_1 \Lambda_1 (ig_s \gamma^{\mu}T^b) v(p_3)\big{\vert}_{q=0}\label{eq.QQ4},
\end{eqnarray}
where $\Lambda_1=\textbf{1}/\sqrt{3}$ is the CS projector.

For the decay channel $Z \to (Q\bar{Q})[^{1}S_0^{[8]}]+Q\bar{Q}$, there are six Feynman diagrams which are shown in Figs.\ref{feyn-QQ-1} and \ref{feyn-QQ-2}. The amplitude (${\cal M}=\sum_{i=1}^{6}{\cal M}_i$) can be written down according to the six Feynman diagrams. The amplitudes ${\cal M}_i(i=1,2,3,4)$ can be obtained from the amplitudes ${\cal M}_i(i=1,2,3,4)$ of $Z \to (Q\bar{Q})[^{1}S_0^{[1]}]+Q\bar{Q}$ through the replacement $\Lambda_1 \to \Lambda_8$. And the remaining two amplitudes ${\cal M}_i(i=5,6)$ are as follows,
\begin{eqnarray}
i {\cal M}_5=&&\frac{i g}{2{\rm cos}\, \theta_W}\frac{-i}{(p_2+p_3)^2+i\epsilon}{\rm tr}\Big[\Pi_{1} \Lambda^a_8\slashed{\epsilon}(p_0) \nonumber  \\
&& \cdot (V_Q-A_Q\gamma_5) \frac{i}{-\slashed{p}_0+\slashed{p}_{11}-m_Q+i \epsilon}  (ig_s \gamma^{\mu}T^b)\Big] \nonumber \\
&& \bar{u}(p_2)(ig_s \gamma_{\mu}T^b)v(p_3)\Big\vert_{q=0}, \label{eq.QQ5}\\
i {\cal M}_6=&&\frac{i g}{2{\rm cos}\, \theta_W}\frac{-i}{(p_2+p_3)^2+i\epsilon}{\rm tr}\Big[\Pi_{1} \Lambda^a_8 (ig_s \gamma^{\mu} T^b)\nonumber  \\
&& \cdot  \frac{i}{\slashed{p}_{0}-\slashed{p}_{12}-m_Q+i \epsilon} \slashed{\epsilon}(p_0) (V_Q-A_Q\gamma_5) \Big]\nonumber \\
&& \bar{u}(p_2)(ig_s \gamma_{\mu}T^b)v(p_3) \Big\vert_{q=0}\label{eq.QQ6}.
\end{eqnarray}
There is an additional factor ($-1$) in ${\cal M}_i(i=5,6)$ compared to ${\cal M}_i(i=1,2,3,4)$, which is due to the fermion exchange.

\begin{figure}[htbp]
\includegraphics[width=0.45\textwidth]{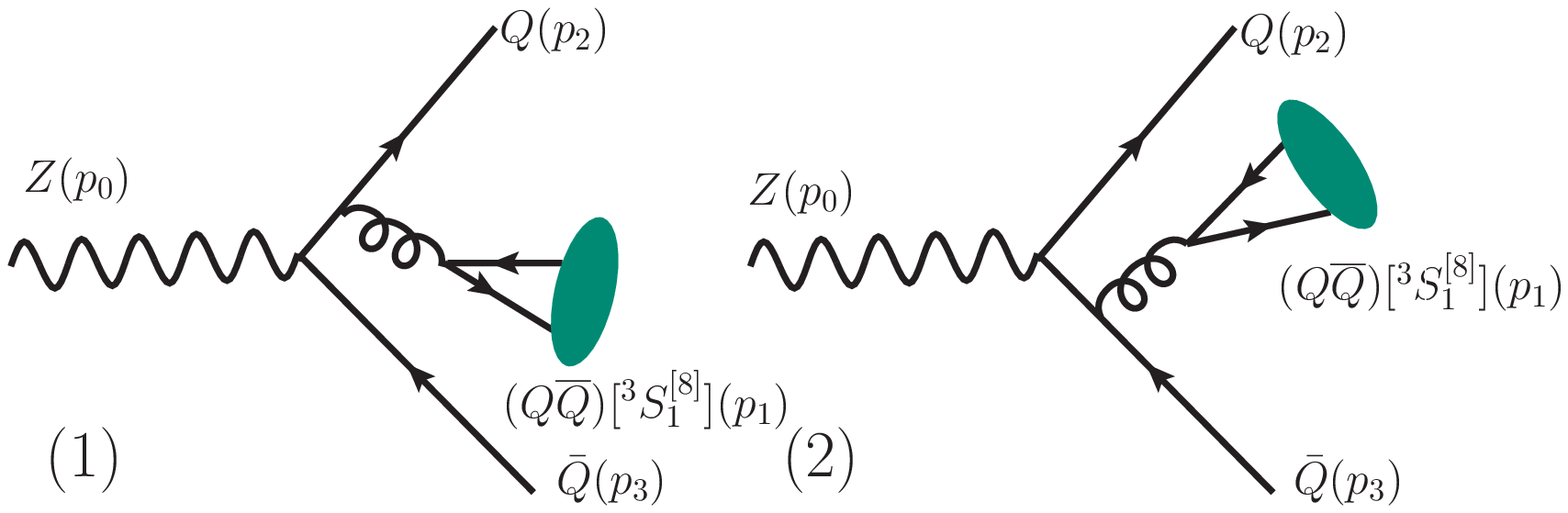}
\caption{Two of the Feynman diagrams for $Z \to (Q\bar{Q})[^{3}S_1^{[8]}]+Q\bar{Q}$.
 } \label{feyn-QQ-3}
\end{figure}

For the decay channel $Z \to (Q\bar{Q})[^{3}S_1^{[8]}]+Q\bar{Q}$, there are eight Feynman diagrams which are shown in Figs.\ref{feyn-QQ-1}, \ref{feyn-QQ-2} and \ref{feyn-QQ-3}. The amplitude can be written as ${\cal M}=\sum_{i=1}^{8}{\cal M}_i$. The amplitudes ${\cal M}_i(i=1,2,3,4)$ can be obtained from Eqs.(\ref{eq.QQ1}), (\ref{eq.QQ2}), (\ref{eq.QQ3}) and (\ref{eq.QQ4}) through the replacement $\Pi_1 \to \Pi_3$ and $\Lambda_1 \to \Lambda_8$. The amplitudes ${\cal M}_i(i=5,6)$ can be obtained from Eqs.(\ref{eq.QQ5}) and (\ref{eq.QQ6}) through the replacement $\Pi_1 \to \Pi_3$. The other two amplitudes ${\cal M}_i(i=7,8)$ are as follows,
\begin{eqnarray}
i {\cal M}_7=&&\frac{i g}{2{\rm cos}\, \theta_W}\frac{-i}{p_1^2+i\epsilon}{\rm tr}\Big[\Pi_3 \Lambda^a_8(ig_s \gamma^{\mu}T^b)\Big] \bar{u}(p_2)(ig_s \gamma_{\mu}T^b) \nonumber  \\
&& \!\!\!\!\!\!\!\!\!\!\!  \cdot \frac{i}{\slashed{p}_1+\slashed{p}_{2}-m_Q+i \epsilon} \slashed{\epsilon}(p_0)(V_Q-A_Q\gamma_5)v(p_3)\Big\vert_{q=0}, \label{eq.QQ7}\\
i {\cal M}_8=&&\frac{i g}{2{\rm cos}\, \theta_W}\frac{-i}{p_1^2+i\epsilon}{\rm tr}\Big[\Pi_3 \Lambda^a_8(ig_s \gamma^{\mu}T^b)\Big] \bar{u}(p_2)\slashed{\epsilon}(p_0) \nonumber  \\
&& \!\!\!\!\!\!\!\!\!\!\!   \cdot (V_Q-A_Q\gamma_5)\frac{i}{-\slashed{p}_1-\slashed{p}_3-m_Q+i \epsilon} (ig_s \gamma_{\mu}T^b) v(p_3)\Big\vert_{q=0}, \label{eq.QQ8}.
\end{eqnarray}

For the decay channel $Z \to (Q\bar{Q})[^{1}P_1^{[8]}]+Q\bar{Q}$, there are six Feynman diagrams which are shown in Figs.\ref{feyn-QQ-1} and \ref{feyn-QQ-2}. The amplitude can be written as ${\cal M}=\sum_{i=1}^{6}{\cal M}_i$, and
\begin{eqnarray}
i{\cal M}_1=&&-\frac{i g\,\epsilon^*_{\alpha}(p_1)}{2{\rm cos}\, \theta_W}\frac{d}{dq_{\alpha}}\Big[\frac{-i}{(p_{12}+p_2)^2+i\epsilon}\bar{u}(p_2)(ig_s \gamma^{\mu}T^b) \nonumber \\
&& \cdot \Pi_1 \Lambda^a_8 (ig_s \gamma_{\mu}T^b) \frac{i}{\slashed{p}_1+\slashed{p}_2-m_Q+i \epsilon} \slashed{\epsilon}(p_0)  \nonumber \\
&& \cdot (V_Q-A_Q\gamma_5)v(p_3)\Big]\Big{\vert}_{q=0}, \label{eq.QQ9} \\
i{\cal M}_2=&&-\frac{i g\,\epsilon^*_{\alpha}(p_1)}{2{\rm cos}\, \theta_W}\frac{d}{dq_{\alpha}}\Big[\frac{-i}{(p_{12}+p_2)^2+i\epsilon}\bar{u}(p_2)(ig_s \gamma^{\mu}T^b) \nonumber \\
&&  \cdot  \Pi_1\Lambda^a_8 \slashed{\epsilon}(p_0)(V_Q-A_Q\gamma_5)\frac{i}{-\slashed{p}_0+\slashed{p}_{11}-m_Q+i \epsilon} \nonumber \\
&&.(ig_s \gamma_{\mu}T^b)v(p_3)\Big]\Big{\vert}_{q=0}, \label{eq.QQ10}\\
i{\cal M}_3=&&-\frac{i g\,\epsilon^*_{\alpha}(p_1)}{2{\rm cos}\, \theta_W}\frac{d}{dq_{\alpha}}\Big[\frac{-i}{(p_{11}+p_3)^2+i\epsilon}\bar{u}(p_2)\slashed{\epsilon}(p_0) \nonumber \\
&&  \cdot (V_Q-A_Q\gamma_5) \frac{i}{-\slashed{p}_1-\slashed{p}_3-m_Q+i \epsilon} (ig_s \gamma_{\mu}T^b) \nonumber \\
&&.\Pi_1 \Lambda^a_8 (ig_s \gamma^{\mu}T^b) v(p_3)\Big]\Big{\vert}_{q=0}, \label{eq.QQ11}\\
i{\cal M}_4=&&-\frac{i g\,\epsilon^*_{\alpha}(p_1)}{2{\rm cos}\, \theta_W}\frac{d}{dq_{\alpha}}\Big[\frac{-i}{(p_{11}+p_3)^2+i\epsilon}\bar{u}(p_2)(ig_s \gamma_{\mu}T^b)  \nonumber \\
&& \cdot \frac{i}{\slashed{p}_0-\slashed{p}_{12}-m_Q+i \epsilon} \slashed{\epsilon}(p_0)(V_Q-A_Q\gamma_5)  \nonumber \\
&&.\Pi_1 \Lambda^a_8 (ig_s \gamma^{\mu}T^b) v(p_3)\Big]\Big{\vert}_{q=0}, \label{eq.QQ12}\\
i {\cal M}_5=&&\frac{i g\,\epsilon^*_{\alpha}(p_1)}{2{\rm cos}\, \theta_W}\frac{-i}{(p_2+p_3)^2+i\epsilon}\frac{d}{dq_{\alpha}}{\rm tr}\Big[\Pi_{1} \Lambda^a_8\slashed{\epsilon}(p_0) \nonumber  \\
&& \cdot (V_Q-A_Q\gamma_5) \frac{i}{-\slashed{p}_0+\slashed{p}_{11}-m_Q+i \epsilon}  (ig_s \gamma^{\mu}T^b)\Big] \nonumber \\
&& \bar{u}(p_2)(ig_s \gamma_{\mu}T^b)v(p_3)\Big\vert_{q=0}, \label{eq.QQ13} \\
i {\cal M}_6=&&\frac{i g\,\epsilon^*_{\alpha}(p_1)}{2{\rm cos}\, \theta_W}\frac{-i}{(p_2+p_3)^2+i\epsilon}\frac{d}{dq_{\alpha}}{\rm tr}\Big[\Pi_{1} \Lambda^a_8 (ig_s \gamma^{\mu} T^b)\nonumber  \\
&& \cdot  \frac{i}{\slashed{p}_{0}-\slashed{p}_{12}-m_Q+i \epsilon} \slashed{\epsilon}(p_0) (V_Q-A_Q\gamma_5)\Big] \nonumber \\
&& \bar{u}(p_2)(ig_s \gamma_{\mu}T^b)v(p_3) \Big\vert_{q=0}. \label{eq.QQ14}
\end{eqnarray}

\subsection{$Z \to \eta_Q(^1S_0^{[8]},^3S_1^{[8]},^1P_1^{[8]})+Q'\bar{Q'}$}

In this subsection we present the amplitudes for the decay channels $Z \to \eta_Q(^1S_0^{[8]},^3S_1^{[8]},^1P_1^{[8]})+Q'\bar{Q'}$ where $Q'$ denotes a heavy quark but $Q'\neq Q$.

\begin{figure}[htbp]
\includegraphics[width=0.45\textwidth]{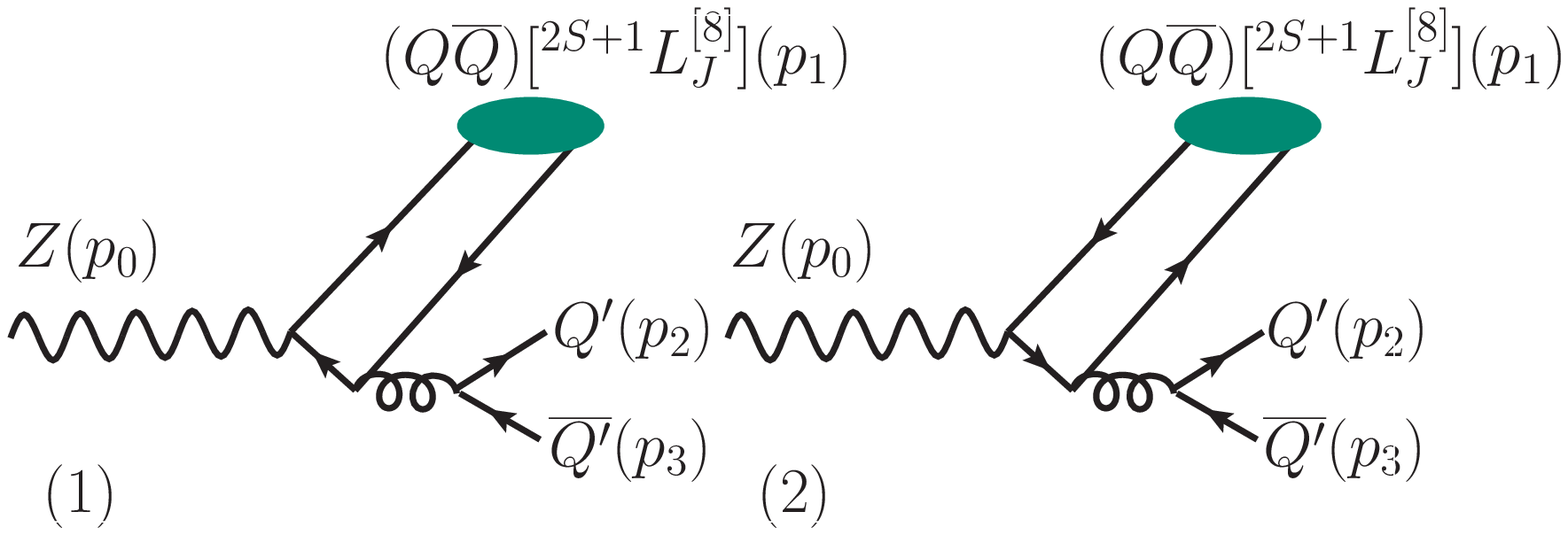}
\caption{Two of the Feynman diagrams for $Z \to (Q\bar{Q})[^{2S+1}L_J^{[8]}]+Q'\bar{Q'}$.
 } \label{feyn-QQ-4}
\end{figure}

For the decay channel $Z \to (Q\bar{Q})[^{1}S_0^{[8]}]+Q'\bar{Q'}$, there are two Feynman diagrams which are shown in Fig.\ref{feyn-QQ-4}. The amplitude can be written as ${\cal M}={\cal M}_1+{\cal M}_2$. The amplitudes ${\cal M}_1$ and ${\cal M}_2$ are the same as ${\cal M}_5$ and ${\cal M}_6$ in Eqs.(\ref{eq.QQ5}) and (\ref{eq.QQ6}), but here we have $p_2^2=p_3^2=m_{Q'}^2$.

\begin{figure}[htbp]
\includegraphics[width=0.45\textwidth]{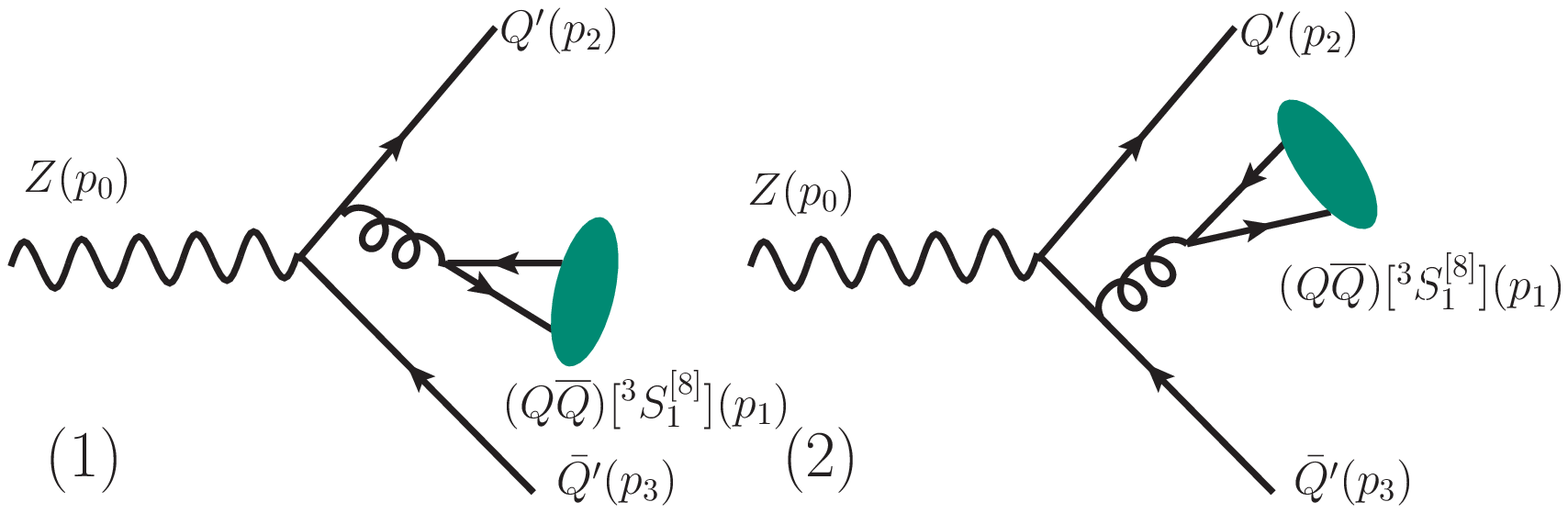}
\caption{Two of the Feynman diagrams for $Z \to (Q\bar{Q})[^{3}S_1^{[8]}]+Q'\bar{Q'}$.
 } \label{feyn-QQ-5}
\end{figure}

For the decay channel $Z \to (Q\bar{Q})[^{3}S_1^{[8]}]+Q'\bar{Q'}$, there are four Feynman diagrams which are shown in Figs.\ref{feyn-QQ-4} and \ref{feyn-QQ-5}. The amplitude can be written as ${\cal M}=\sum_{i=1}^4 {\cal M}_i$. The amplitudes ${\cal M}_1$ and ${\cal M}_2$ can be obtained from ${\cal M}_5$ and ${\cal M}_6$ in Eqs.(\ref{eq.QQ5}) and (\ref{eq.QQ6}) through replacement $\Pi_1 \to \Pi_3$. The amplitudes ${\cal M}_3$ and ${\cal M}_4$, which correspond to the two diagrams in Fig.\ref{feyn-QQ-5}, are as follows:
\begin{eqnarray}
i {\cal M}_3=&&\frac{i g}{2{\rm cos}\, \theta_W}\frac{-i}{p_1^2+i\epsilon}{\rm tr}\Big[\Pi_3 \Lambda^a_8(ig_s \gamma^{\mu}T^b)\Big] \bar{u}(p_2)(ig_s \gamma_{\mu}T^b) \nonumber  \\
&&\!\!\!\!\!\!\!\!\!\!\!  \cdot \frac{i}{\slashed{p}_1+\slashed{p}_{2}-m_{Q'}+i \epsilon} \slashed{\epsilon}(p_0)(V_{Q'}-A_{Q'}\gamma_5)v(p_3)\Big\vert_{q=0},\\
i {\cal M}_4=&&\frac{i g}{2{\rm cos}\, \theta_W}\frac{-i}{p_1^2+i\epsilon}{\rm tr}\Big[\Pi_3 \Lambda^a_8(ig_s \gamma^{\mu}T^b)\Big] \bar{u}(p_2)\slashed{\epsilon}(p_0) \nonumber  \\
&&\!\!\!\!\!\!\!\!\!\!\!  \cdot (V_{Q'}-A_{Q'}\gamma_5)\frac{i}{-\slashed{p}_1-\slashed{p}_3-m_{Q'}+i \epsilon} (ig_s \gamma_{\mu}T^b) v(p_3)\Big\vert_{q=0}.
\end{eqnarray}

For the decay channel $Z \to (Q\bar{Q})[^{1}P_1^{[8]}]+Q'\bar{Q'}$, there are two Feynman diagrams which are shown in Fig.\ref{feyn-QQ-4}. The amplitude can be written as ${\cal M}={\cal M}_1+{\cal M}_2$. The amplitudes ${\cal M}_1$ and ${\cal M}_2$ have the same form as ${\cal M}_5$ and ${\cal M}_6$ in Eqs.(\ref{eq.QQ13}) and (\ref{eq.QQ14}), but here $p_2^2=p_3^2=m_{Q'}^2$.

\subsection{$Z \to \eta_Q(^1S_0^{[1]})+gg$}

\begin{figure}[htbp]
\includegraphics[width=0.45\textwidth]{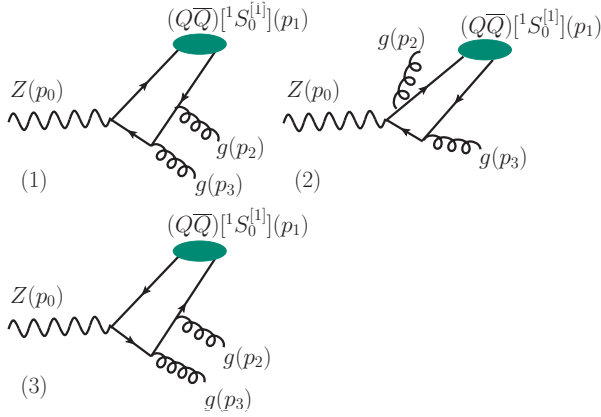}
\caption{Half of the Feynman diagrams for $Z \to (Q\bar{Q})[^{1}S_0^{[1]}]+gg$.
 } \label{feyn-gg}
\end{figure}

There are six Feynman diagrams for the decay channel $Z \to (Q\bar{Q})[^1S_0^{[1]}] +gg$. Half of the Feynman diagrams are shown in Fig.\ref{feyn-gg}, and the other three Feynman diagrams can be obtained from these diagrams through the substitution $p_2 \leftrightarrow p_3$. According to those diagrams, the amplitude (${\cal M}=\sum_{i=1}^{6} {\cal M}_i$) of the process can be written down, and we have
\begin{eqnarray}
i {\cal M}_1=&&-\frac{i g}{2{\rm cos}\, \theta_W}{\rm tr}\Big[\Pi_1 \Lambda_1 \slashed{\epsilon}(p_0)(V_Q-A_Q\gamma_5) \nonumber \\
&&\cdot \frac{i}{-\slashed{p}_0+\slashed{p}_{11}-m_Q+i \epsilon}  (ig_s \slashed{\epsilon}^*(p_3)T^b)\nonumber \\
&&\cdot \frac{i}{-\slashed{p}_2-\slashed{p}_{12}-m_Q+i \epsilon} (ig_s \slashed{\epsilon}^*(p_2)T^a)\Big]\Big\vert_{q=0}, \\
i {\cal M}_2=&&-\frac{i g}{2{\rm cos}\, \theta_W}{\rm tr}\Big[\Pi_1 \Lambda_1 (ig_s \slashed{\epsilon}^*(p_2) T^a) \nonumber \\
&&\cdot \frac{i}{\slashed{p}_2+\slashed{p}_{11}-m_Q+i \epsilon} \slashed{\epsilon}(p_0) (V_Q-A_Q\gamma_5)\nonumber \\
&&\cdot \frac{i}{-\slashed{p}_3-\slashed{p}_{12}-m_Q+i \epsilon} (ig_s \slashed{\epsilon}^*(p_3)T^b)\Big]\Big\vert_{q=0},\\
i {\cal M}_3=&&-\frac{i g}{2{\rm cos}\, \theta_W}{\rm tr}\Big[\Pi_1 \Lambda_1  (ig_s \slashed{\epsilon}^*(p_2)T^a) \nonumber \\
&&\cdot \frac{i}{\slashed{p}_2+\slashed{p}_{11}-m_Q+i \epsilon}  (ig_s \slashed{\epsilon}^*(p_3)T^b)\nonumber \\
&&\cdot \frac{i}{\slashed{p}_0-\slashed{p}_{12}-m_Q+i \epsilon}\slashed{\epsilon}(p_0) (V_Q-A_Q\gamma_5)\Big]\Big\vert_{q=0}.
\end{eqnarray}
The other three amplitudes $M_i(i=4,5,6)$ can be obtained from $M_i(i=1,2,3)$ via the substitution $p_2 \leftrightarrow p_3$.

\section{Numerical results}
\label{secNumer}

In the calculations, the package FeynArts \cite{feynarts} is employed to generate Feynman diagrams and amplitudes, the package FeynCalc \cite{feyncalc1,feyncalc2} is employed to carry out the color and Dirac traces, the package \$Apart \cite{apart} is employed to conduct the partial fraction, the package FIRE \cite{fire} is employed to do the integration-by-parts (IBP) reduction, and the package LoopTools \cite{looptools} is used to compute the one-loop master integrals numerically. The phase-space integrations are performed by using the package Vegas \cite{vegas}.

The necessary input parameters for the numerical calculation are taken as follows:
\begin{eqnarray}
&& m_c=1.5\,{\rm GeV},\; m_b=4.75\,{\rm GeV},\; m_{_Z}=91.1876\,{\rm GeV},\nonumber \\
&& {\rm sin}^2\theta_W=0.231,\alpha=1/128,
\end{eqnarray}
where $\alpha$ is the electromagnetic coupling constant at $m_{_Z}$. For the strong coupling constant, we adopt the one-loop formula
\begin{displaymath}
\alpha_s(\mu_R)=\frac{4\pi}{\beta_0\,{\rm ln}(\mu_R^2/\Lambda^2_{QCD})}.
\end{displaymath}
According to $\alpha_s(m_{_Z})=0.1179$~\cite{pdg}, we obtain $\alpha_s(2m_c)=0.234$ and $\alpha_s(2m_b)=0.175$.

For the LDMEs, we derive the LDMEs for the $\eta_Q$ from  the experimentally extracted LDMEs for the $J/\psi(\Upsilon)$ via the heavy quark spin symmetry (HQSS), i.e.,
\begin{eqnarray}
\langle {\cal O}^{\eta_Q}(^{1}S_{0}^{[1]}/^{1}S_{0}^{[8]})\rangle &=& \frac{1}{3}\langle {\cal O}^{\psi_Q}(^{3}S_{1}^{[1]}/^{3}S_{1}^{[8]})\rangle,\nonumber \\
\langle {\cal O}^{\eta_Q}(^{3}S_{1}^{[8]})\rangle &=& \langle {\cal O}^{\psi_Q}(^{1}S_{0}^{[8]})\rangle,\nonumber \\
\langle {\cal O}^{\eta_Q}(^{1}P_{1}^{[8]})\rangle &=& 3\, \langle {\cal O}^{\psi_Q}(^{3}P_{0}^{[8]})\rangle.
\end{eqnarray}
These relations are expected to hold to relative order $v^2_Q$. Several sets of the LDMEs for the $J/\psi$ and the $\Upsilon$ extracted from the global fits by several groups are listed in Tables \ref{tb.LDMEs-1} and \ref{tb.LDMEs-2}. The factorization scale of the LDMEs has been taken as $\mu_{\Lambda}=1.5\,{\rm GeV}$ for the $J/\psi$ and the $\Upsilon$ in Tables \ref{tb.LDMEs-1} and \ref{tb.LDMEs-2}. Thus, we also take this value for $\mu_{\Lambda}$ in this paper.

\begin{table}[htb]
\begin{tabular}{c c c c}
\hline
\multirow{2}{*}{LDMEs}
 &  Butenschoen & Chao & Gong \\
 & et al. \cite{Butenschoen:2011yh} & et al. \cite{Chao:2012iv} & et al. \cite{Gong:2012ug} \\
\hline
$\langle {\cal O}^{J/\psi}(^{3}S_{1}^{[1]})\rangle/{\rm GeV}^3$  & 1.32  & 1.16 &  1.16  \\
$\langle {\cal O}^{J/\psi}(^{1}S_{0}^{[8]})\rangle/(10^{-2}{\rm GeV}^3)$  & 3.04   & 8.9  & 9.7  \\
$\langle {\cal O}^{J/\psi}(^{3}S_{1}^{[8]})\rangle/(10^{-2}{\rm GeV}^3)$ & 0.17 &  0.30 & -0.46   \\
$\langle {\cal O}^{J/\psi}(^{3}P_{0}^{[8]})\rangle/(10^{-2}{\rm GeV}^5)$ &  -0.91 & 1.26 & -2.14   \\
\hline
\end{tabular}
\caption{The LDMEs for the $J/\psi$ production extracted from the global fits, where $\mu_{\Lambda}=m_c$.}
\label{tb.LDMEs-1}
\end{table}

\begin{table}[htb]
\begin{tabular}{c c c}
\hline
\multirow{2}{*}{LDMEs}
  &  Gong & Feng  \\
  & et al. \cite{Gong:2013qka} & et al. \cite{Feng:2015wka} \\
\hline
$\langle {\cal O}^{\Upsilon}(^{3}S_{1}^{[1]})\rangle/{\rm GeV}^3$  & 9.28  & 9.28 \\
$\langle {\cal O}^{\Upsilon}(^{1}S_{0}^{[8]})\rangle/(10^{-2}{\rm GeV}^3)$  &  11.15 & 13.6  \\
$\langle {\cal O}^{\Upsilon}(^{3}S_{1}^{[8]})\rangle/(10^{-2}{\rm GeV}^3)$ & -0.41 &    0.61  \\
$\langle {\cal O}^{\Upsilon}(^{3}P_{0}^{[8]})\rangle/m_b^2/(10^{-2}{\rm GeV}^3)$ & -0.67  & -0.93  \\
\hline
\end{tabular}
\caption{The LDMEs for the $\Upsilon$ production extracted from the global fits, where $\mu_{\Lambda}=1.5\,{\rm GeV}$.}
\label{tb.LDMEs-2}
\end{table}

\subsection{Integrated decay widths}

In this subsection, we give the decay widths for different decay channels and the total decay widths for the inclusive $\eta_Q$ production via $Z$ boson decays.

\begin{table}[htb]
\begin{tabular}{l c c c}
\hline
 Decay &  Butenschoen & Chao & Gong \\
 channels & et al.  & et al. & et al. \\
\hline
$\eta_c(^1S_0^{[8]})+g$\,(LO)  &  $3.29\times 10^{-3}$ & $5.80\times 10^{-3}$ &  $-8.90\times 10^{-3}$  \\
$\eta_c(^1S_0^{[8]})+g$\,(NLO)  &  $1.79\times 10^{-2}$ & $3.15\times 10^{-2}$ &  $-4.83\times 10^{-2}$  \\
$\eta_c(^3S_1^{[8]})+g$\,(LO)  &  0.399   & 1.17  & 1.27   \\
$\eta_c(^3S_1^{[8]})+g$\,(NLO)  &  396   & $1.16\times 10^{3}$  & $1.26\times 10^{3}$   \\
$\eta_c(^1P_1^{[8]})+g$\,(LO) & -0.159  & 0.221  &  -0.375  \\
$\eta_c(^1P_1^{[8]})+g$\,(NLO) & -0.263  & 0.364  &  -0.618 \\
Total($\eta_c+g$)\,(LO) &  0.243   &  1.40 &  0.886  \\
Total($\eta_c+g$)\,(NLO) &  396   &  $1.16\times 10^{3}$ &  $1.26\times 10^{3}$  \\
\hline
\end{tabular}
\caption{The decay widths (unit:keV) for the decay channels  $Z \to \eta_c(^{2S+1}L_J^{[8]})+g$ based on three sets of LDMEs, where ``NLO" denotes the results up to NLO accuracy. The very large NLO correction in the $^3S_1^{[8]}$ case comes from the contribution of the real correction processes $Z \to \eta_c(^{3}S_1^{[8]})+q\bar{q}$.}
\label{tb.Zetac-g}
\end{table}

\begin{table}[htb]
\begin{tabular}{c c c c}
\hline
 Decay &  Butenschoen & Chao & Gong \\
 channels & et al. & et al. & et al. \\
\hline
$\eta_c(^1S_0^{[1]})+c\bar{c}$  &  90.9 & 79.9 &  79.9  \\
$\eta_c(^1S_0^{[8]})+c\bar{c}$  &  $1.10\times 10^{-2}$ & $1.94\times 10^{-2}$ &  $-2.98\times 10^{-2}$  \\
$\eta_c(^3S_1^{[8]}])+c\bar{c}$  & 104   &  305 &  332  \\
$\eta_c(^1P_1^{[8]}])+c\bar{c}$ &  $-8.84\times 10^{-2}$ & 0.122 &   -0.208 \\
 Total($\eta_c+c\bar{c}$) &  195  & 385  &  412  \\
\hline
\end{tabular}
\caption{The decay widths (unit:keV) for the decay channels  $Z \to \eta_c(^{2S+1}L_J^{[1,8]})+c\bar{c}$ based on three sets of LDMEs.}
\label{tb.Zetac-cc}
\end{table}

\begin{table}[htb]
\begin{tabular}{c c c c}
\hline
 Decay &  Butenschoen & Chao & Gong \\
 channels & et al. & et al. & et al.  \\
\hline
$\eta_c(^1S_0^{[8]})+b\bar{b}$  & $8.53\times 10^{-5}$ & $1.51\times 10^{-4}$ & $-2.31\times 10^{-4}$   \\
$\eta_c(^3S_1^{[8]}])+b\bar{b}$  & 117   & 342  & 373  \\
$\eta_c(^1P_1^{[8]}])+b\bar{b}$ &  $-3.15\times 10^{-3}$ & $4.36\times 10^{-3}$ &   $-7.40\times 10^{-3}$ \\
 Total($\eta_c+b\bar{b}$) & 117   & 342  &  373  \\
\hline
\end{tabular}
\caption{The decay widths (unit:keV) for the decay channels  $Z \to \eta_c(^{2S+1}L_J^{[8]})+b\bar{b}$ based on three sets of LDMEs.}
\label{tb.Zetac-bb}
\end{table}

\begin{table}[htb]
\begin{tabular}{c c c c}
\hline
 Decay &  Butenschoen & Chao & Gong \\
 channels & et al.  & et al. & et al.  \\
\hline
$\eta_c(^1S_0^{[1]})+gg$  &  4.43  & 3.89  &  3.89  \\
\hline
\end{tabular}
\caption{The decay width (unit:keV) for the decay channel  $Z \to \eta_c(^{1}S_0^{[1]})+gg$ based on three sets of LDMEs.}
\label{tb.Zetac-gg}
\end{table}

The decay widths for the decay channels contributing to $Z \to \eta_c+X$ are given in Tables \ref{tb.Zetac-g}, \ref{tb.Zetac-cc}, \ref{tb.Zetac-bb} and \ref{tb.Zetac-gg}.  In Table \ref{tb.Zetac-g}, the decay widths for the decay channels $Z \to \eta_c(^{2S+1}L_J^{[8]})+g$ up to LO and NLO accuracy in $\alpha_s$ are presented. We can see that the NLO correction is larger than the LO contribution in the $^1S_0^{[8]}$ and $^3S_1^{[8]}$ cases. The reason of the large NLO correction in the $^1S_0^{[8]}$ case is that only the vector coupling of the $Z-c\bar{c}$ vertex contributes to the decay width of $Z \to \eta_c(^{1}S_0^{[8]})+g$ at the LO level, while both the vector and axial-vector couplings of the $Z-c\bar{c}$ vertex contribute to the NLO correction through the real corrections. Moreover, the strength of the axial-vector coupling is stronger than that of the vector coupling in the $Z-c\bar{c}$ vertex. Thus, the large NLO correction in the $^1S_0^{[8]}$ case is expected. The reason for the very large NLO correction in the $^3S_1^{[8]}$ case is that there are gluon fragmentation diagrams for the processes $Z \to \eta_c(^{3}S_1^{[8]})+q\bar{q}$ at $\alpha \alpha_s^2$ order. One of the gluon fragmentation diagram is the sixth diagram in Fig.\ref{feyn-g-real}. The decay width for $Z \to \eta_c(^{3}S_1^{[8]})+q\bar{q}(q=u,d,s)$ \footnote{When we present the results for $Z \to \eta_Q(^{3}S_1^{[8]})+q\bar{q}$ individually, we actually give the contribution from the fragmentation diagrams, which is gauge invariant and counts almost the whole contribution of the NLO decay width of $Z\to \eta_c(^{3}S_1^{[8]})+g$.} is $1.15\times 10^3 \,{\rm keV}$ under the LDME extracted by Chao et al, which is very close to the decay width for $Z \to \eta_c(^{3}S_1^{[8]})+g$ at the NLO level.

From these tables, we can see that the dominant contributions come from the decay channels $Z \to \eta_c(^1S_0^{[1]},^3S_1^{[8]})+c\bar{c}$, $Z \to \eta_c(^3S_1^{[8]})+b\bar{b}$ and $Z \to \eta_c(^3S_1^{[8]})+q\bar{q}$. Among these dominant channels, the CO ($^3S_1^{[8]}$) channels are more important than the CS ($^1S_0^{[1]}$) channel under the three sets of LDMEs.
The LO contributions from the decay channels associated with a final gluon are suppressed although they are of order $\alpha \alpha_s$.

These dominant decay channels can be understood by the fragmentation mechanism. For the decay channel $Z \to \eta_c(^1S_0^{[1]})+c\bar{c}$, the decay width is dominated by the (anti)quark fragmentation process of $Z \to c\bar{c}$ followed by $c(\bar{c})\to \eta_c$. In this fragmentation process, the quark propagator is of order $1/m_{\eta_c}$, and the gluon propagator is of order $1/m_{\eta_c}^2$. For the decay channels $Z \to \eta_c(^1S_0^{[8]},^1P_1^{[8]})+c\bar{c}$, the decay widths are also dominated by the (anti)quark fragmentation process. However, the involved LDMEs in the two decay channels are suppressed by powers of $v_c$ compared with the CS LDME. Thus, the decay widths of the decay channels $Z \to \eta_c(^1S_0^{[8]},^1P_1^{[8]})+c\bar{c}$ are suppressed compared with that of $Z \to \eta_c(^1S_0^{[1]})+c\bar{c}$. For the decay channels $Z \to \eta_c(^3S_1^{[8]})+c\bar{c}(b\bar{b},q\bar{q})$, the decay widths are dominated by the fragmentation processes of $Z \to c\bar{c}(b\bar{b},q\bar{q})$ followed by a quark or an antiquark fragments into the $\eta_c$ and $Z \to c\bar{c}g(b\bar{b}g,q\bar{q}g)$ followed by $g \to \eta_c$. In the fragmentation processes $Z \to c\bar{c}(b\bar{b},q\bar{q})$ followed by $c(b,q,\bar{c},\bar{b},\bar{q}) \to \eta_c$, the quark propagator is of order $1/m_{\eta_c}$, and the gluon propagator is $1/m_{\eta_c}^2$. Since the gluon propagator is fixed as $1/m_{\eta_c}^2$ in the whole phase space, the CO channels $Z \to \eta_c(^3S_1^{[8]})+c\bar{c}(b\bar{b},q\bar{q})$ are more important than the CS channel $Z \to \eta_c(^1S_0^{[1]})+c\bar{c}$ although the CO channels are suppressed by powers of $v_c$ compared with the CS channel. The decay channels $Z \to \eta_c(^1S_0^{[8]}, ^3S_1^{[8]}, ^1P_1^{[8]}) +g$ at LO in $\alpha_s$ have no fragmentation contribution, the quark propagator in these channels is of order $1/m_{_Z}$. Other channels $Z \to \eta_c(^1S_0^{[8]},^1P_1^{[8]}) +b\bar{b}$ and $Z \to \eta_c(^1S_0^{[1]}) +gg$ also have no fragmentation contribution, thus they are suppressed. \footnote{Actually, the decay widths for the heavy quarkonium production can be further organized by different powers of $m_Q/m_{_Z}$ under the fragmentation-function approach, more detailed discussions for the power expansion can be found in Refs.\cite{Chang:1994aw, Kang:2011zza, Kang:2011mg, Kang:2014tta, Lee:2020dza, Fleming:2012wy}.}

Due to the fact that the SDCs of the $^3S_1^{[8]}$ channels are greatly enhanced compared to other channels and the $^3S_1^{[8]}$ channels dominate the decay $Z\to \eta_c+X$, the total decay width of $Z\to \eta_c+X$ is sensitive to the CO LDME $\langle {\cal O}^{\eta_c}(^{3}S_{1}^{[8]})\rangle$. Therefore, the process $Z\to \eta_c+X$ provides a good platform to determine the value of $\langle {\cal O}^{\eta_c}(^{3}S_{1}^{[8]})\rangle$. Moreover, according to HQSS, we have $\langle {\cal O}^{J/\psi}(^{1}S_{0}^{[8]})\rangle=\langle {\cal O}^{\eta_c}(^{3}S_{1}^{[8]})\rangle(1+{\cal O}(v_c^2))$. The value of $\langle {\cal O}^{\eta_c}(^{3}S_{1}^{[8]})\rangle$ can give a good constraint to the value of $\langle {\cal O}^{J/\psi}(^{1}S_{0}^{[8]})\rangle$.

\begin{table}[htb]
\begin{tabular}{c c c c}
\hline
&  Butenschoen et al.& Chao et al. & Gong et al.\\
\hline
$\eta_c+X$  & 0.712 & 1.89 &   2.05 \\
\hline
\end{tabular}
\caption{The decay width (unit:MeV) for $Z \to \eta_c+X$ based on three sets of LDMEs.}
\label{tb.Zetac-X}
\end{table}

Summing the contributions from the considered decay channels, we obtain the decay width for the inclusive process $Z \to \eta_c +X$ which is given in Table \ref{tb.Zetac-X}.

\begin{table}[htb]
\begin{tabular}{c c c}
\hline
 Decay channels&  Gong et al. & Feng et al. \\
\hline
$\eta_b(^1S_0^{[8]})+g$(LO)  & $-6.02 \times 10^{-3}$  & $8.96 \times 10^{-3}$   \\
$\eta_b(^1S_0^{[8]})+g$(NLO)  & $-1.13 \times 10^{-2}$  & $1.68 \times 10^{-2}$   \\
$\eta_b(^3S_1^{[8]})+g$(LO)  &   0.346  & 0.422   \\
$\eta_b(^3S_1^{[8]})+g$(NLO)  &   8.27  & 10.1   \\
$\eta_b(^1P_1^{[8]})+g$(LO) & $-6.23 \times 10^{-2}$  & $-8.65 \times 10^{-2}$    \\
$\eta_b(^1P_1^{[8]})+g$(NLO) & -0.107  & -0.149    \\
Total($\eta_b+g$)(LO) &  0.278   &  0.344  \\
Total($\eta_b+g$)(NLO) & 8.15    &  9.97  \\
\hline
\end{tabular}
\caption{The decay widths (unit:keV) for the decay channels  $Z \to \eta_b(^{2S+1}L_J^{[8]})+g$ based on two sets of LDMEs, where ``NLO" denotes the results up to NLO accuracy. The very large NLO correction in the $^3S_1^{[8]}$ case comes from the contribution of the real correction processes $Z \to \eta_b(^{3}S_1^{[8]})+q\bar{q}$.}
\label{tb.Zetab-g}
\end{table}

\begin{table}[htb]
\begin{tabular}{c c c}
\hline
Decay channels&  Gong et al. & Feng et al. \\
\hline
$\eta_b(^1S_0^{[1]})+b\bar{b}$  & 10.5  &  10.5  \\
$\eta_b(^1S_0^{[8]})+b\bar{b}$  & $-4.65 \times 10^{-4}$  & $6.92 \times 10^{-4}$ \\
$\eta_b(^3S_1^{[8]}])+b\bar{b}$  &  2.49  & 3.04   \\
$\eta_b(^1P_1^{[8]}])+b\bar{b}$ &  $-2.88 \times 10^{-3}$  &  $-3.99 \times 10^{-3}$   \\
 Total($\eta_b+b\bar{b}$) &  13.0  &   13.5  \\
\hline
\end{tabular}
\caption{The decay widths (unit:keV) for the decay channels $Z \to \eta_b(^{2S+1}L_J^{[1,8]})+b\bar{b}$ based on two sets of LDMEs.}
\label{tb.Zetab-bb}
\end{table}

\begin{table}[htb]
\begin{tabular}{c c c}
\hline
Decay channels&  Gong et al. & Feng et al. \\
\hline
$\eta_b(^1S_0^{[8]})+c\bar{c}$  & $-2.40 \times 10^{-4}$ & $3.58 \times 10^{-4}$ \\
$\eta_b(^3S_1^{[8]}])+c\bar{c}$  & 2.28   & 2.78  \\
$\eta_b(^1P_1^{[8]}])+c\bar{c}$ & $-2.18 \times 10^{-3}$  & $-3.02 \times 10^{-3}$ \\
 Total($\eta_b+c\bar{c}$) &  2.28  &  2.78  \\
\hline
\end{tabular}
\caption{The decay widths (unit:keV) for the decay channels  $Z \to \eta_b(^{2S+1}L_J^{[8]})+c\bar{c}$ based on two sets of LDMEs.}
\label{tb.Zetab-cc}
\end{table}

\begin{table}[htb]
\begin{tabular}{c c c}
\hline
 Decay channels&  Gong et al. & Feng et al. \\
\hline
$\eta_b(^1S_0^{[1]})+gg$  &  2.10  &  2.10   \\
\hline
\end{tabular}
\caption{The decay width (unit:keV) for the decay channel  $Z \to \eta_b(^{1}S_0^{[1]})+gg$ based on two sets of LDMEs.}
\label{tb.Zetab-gg}
\end{table}

The contributions to the decay width of $Z \to \eta_b+X$ from the considered decay channels are given in Tables \ref{tb.Zetab-g}, \ref{tb.Zetab-bb}, \ref{tb.Zetab-cc} and \ref{tb.Zetab-gg}. The very large NLO correction to the decay channel $Z \to \eta_b(^{3}S_1^{[8]})+g$ comes from the processes $Z \to \eta_b(^{3}S_1^{[8]})+q\bar{q}$ whose decay width is $8.21\,{\rm keV}$ under the LDME extracted by Gong et al. Similar to the $\eta_c$ case, the dominant contributions come from the decay channels  $Z \to \eta_b(^1S_0^{[1]},^3S_1^{[8]})+b\bar{b}$, $Z \to \eta_b(^3S_1^{[8]})+c\bar{c}$ and $Z \to \eta_b(^3S_1^{[8]})+q\bar{q}$ due to the fragmentation mechanism in these channels.

Different from the $\eta_c$ case, among these dominant decay channels, the CS channel $Z \to \eta_b(^1S_0^{[1]})+b\bar{b}$ is the most important channel. There are two reasons: One is that $m_{_Z}/m_{\eta_b}$ is smaller than $m_{_Z}/m_{\eta_c}$, which leads to the enhancement in the SDCs of the $^3S_1^{[8]}$ channels is weakened for the $\eta_b$ case; The other is that the CO LDME $\langle {\cal O}^{\eta_Q}(^{3}S_{1}^{[8]})\rangle$ is more suppressed compared with the CS LDME $\langle {\cal O}^{\eta_Q}(^{1}S_{0}^{[1]})\rangle$ in the $\eta_b$ case. However, for the $^{3}S_1^{[8]}$ channels, since the final open quark pair can be several flavors, the sum of these $^{3}S_1^{[8]}$ channels dominate the decay $Z \to \eta_b+X$. Therefore, the decay $Z \to \eta_b+X$ can be used to determine the value of $\langle {\cal O}^{\eta_b}(^{3}S_{1}^{[8]})\rangle$, and give a good constraint to the value of $\langle {\cal O}^{\Upsilon}(^{1}S_{0}^{[8]})\rangle$.

\begin{table}[htb]
\begin{tabular}{c c c}
\hline
  &  Gong et al. & Feng et al. \\
\hline
$\eta_b+X$  & 25.5 &  28.4\\
\hline
\end{tabular}
\caption{The decay width (unit:keV) for $Z \to \eta_b+X$ based on two sets of LDMEs.}
\label{tb.Zetab-X}
\end{table}

Summing the contributions from the considered decay channels, we obtain the decay width for the inclusive process $Z \to \eta_b +X$ which is presented in Table \ref{tb.Zetab-X}.

\subsection{Differential decay widths}

\begin{figure}[htb]
\includegraphics[width=0.45\textwidth]{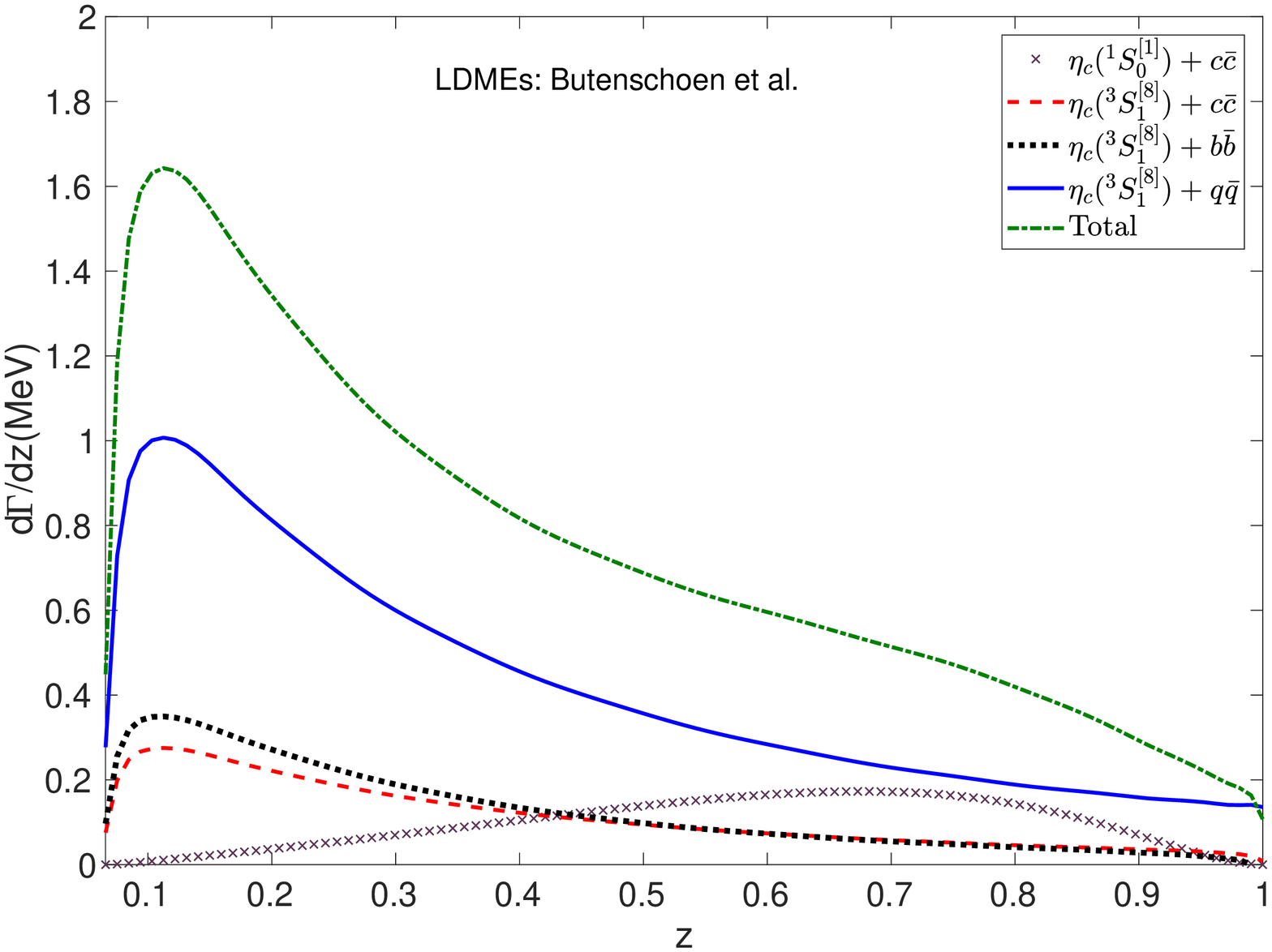}
\includegraphics[width=0.45\textwidth]{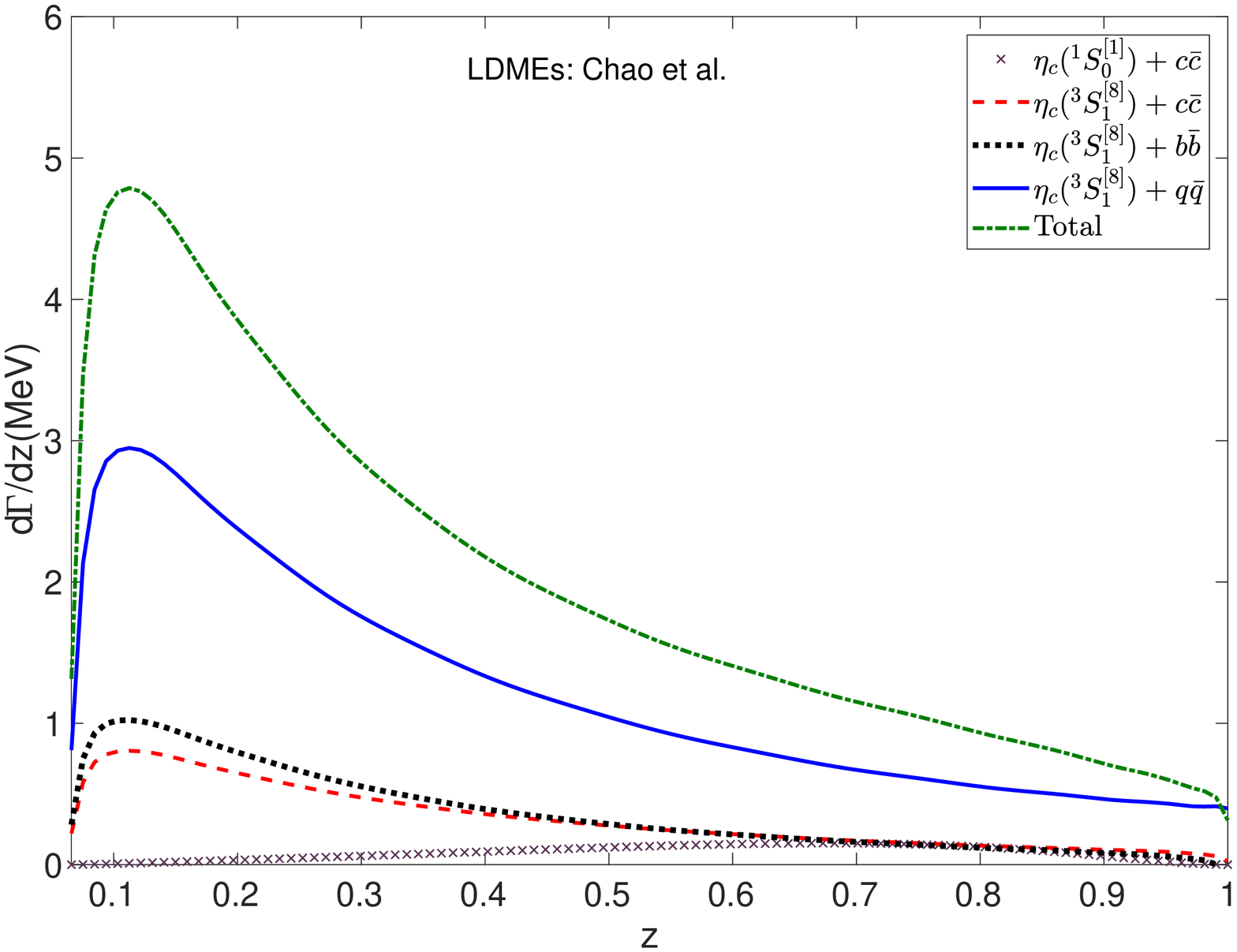}
\includegraphics[width=0.45\textwidth]{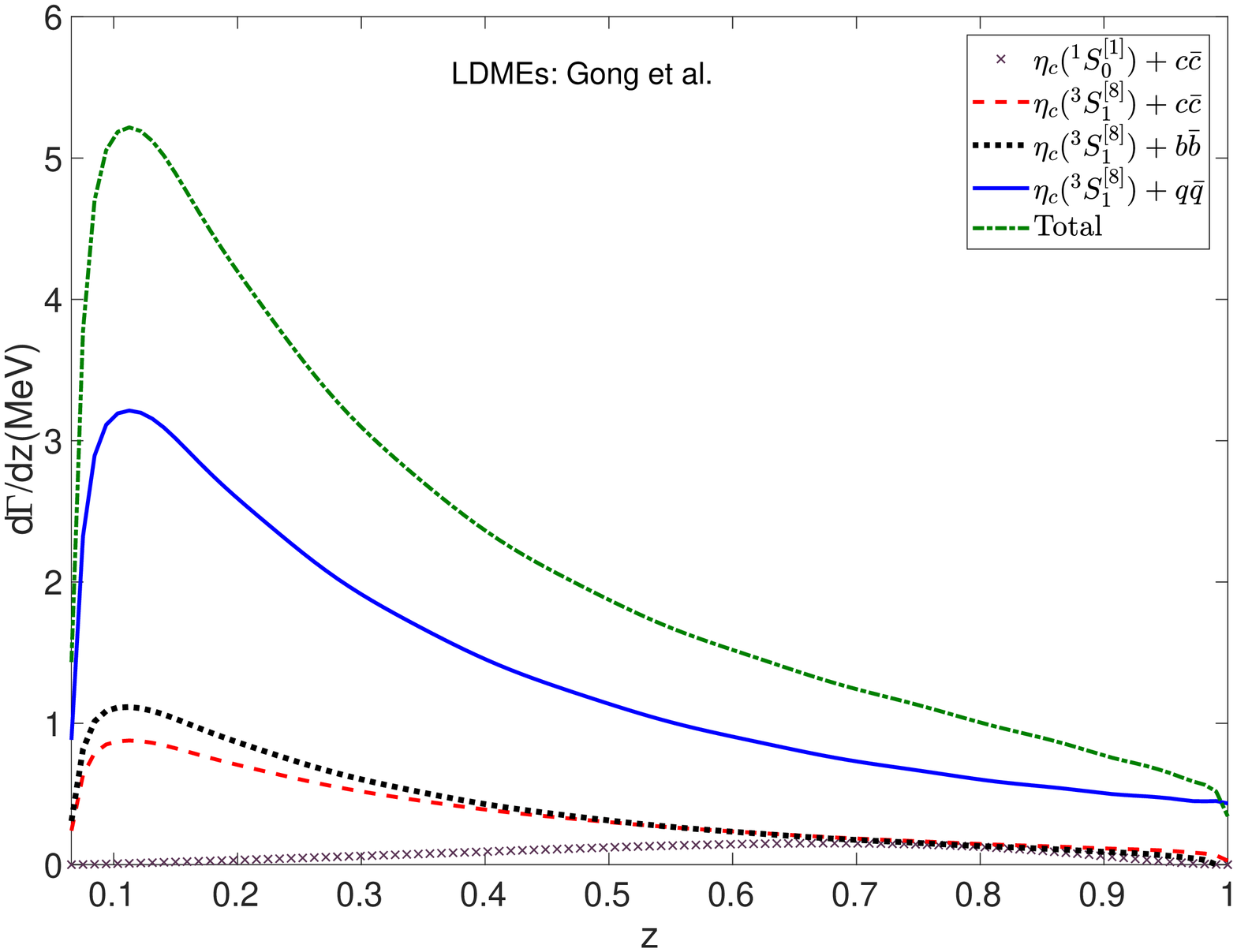}
\caption{The differential decay widths $d\Gamma/dz$ for $Z \to \eta_c+X$, where $\mu_R=2m_c$. The top one shows the distributions based on the LDMEs of Butenschoen et al \cite{Butenschoen:2011yh}, the middle one shows the distributions based the LDMEs of Chao et al \cite{Chao:2012iv}, and the bottom one shows the distributions based the LDMEs of Gong et al \cite{Gong:2012ug}.
 } \label{gammazc}
\end{figure}

In this subsection, we present the differential decay widths $d\Gamma/dz$ for $Z \to \eta_Q+X$, where the energy fraction is defined as $z\equiv 2p_{\eta_Q}\cdot p_{_Z}/p_{_Z}^2$. Since the dominant contributions to the decay process $Z \to \eta_Q+X$ come from the $^1S_0^{[1]}$ and $^3S_1^{[8]}$ channels and the contributions from other channels are greatly suppressed, we only consider the $^1S_0^{[1]}$ and $^3S_1^{[8]}$ channels in this subsection.

The differential decay widths $d\Gamma/dz$ for $Z \to \eta_c+X$ based on the three sets of LDMEs are given in Fig.\ref{gammazc}. From the figure, we can see that the $^3S_1^{[8]}$ channels dominate the decay $Z \to \eta_c +X$, which was also shown in the last subsection by the integrated decay widths. The distributions of the CS channel and the CO channels have different shapes. The curve of the CS channel has a peak at a moderate $z$ value, while the curves of the CO channels have a peak at a small $z$ value. This feature can be used to determine the CS LDME $\langle {\cal O}^{\eta_c}(^{1}S_{0}^{[1]})\rangle$ and the CO LDME $\langle {\cal O}^{\eta_c}(^{3}S_{1}^{[8]})\rangle$ more precisely.

\begin{figure}[htb]
\includegraphics[width=0.45\textwidth]{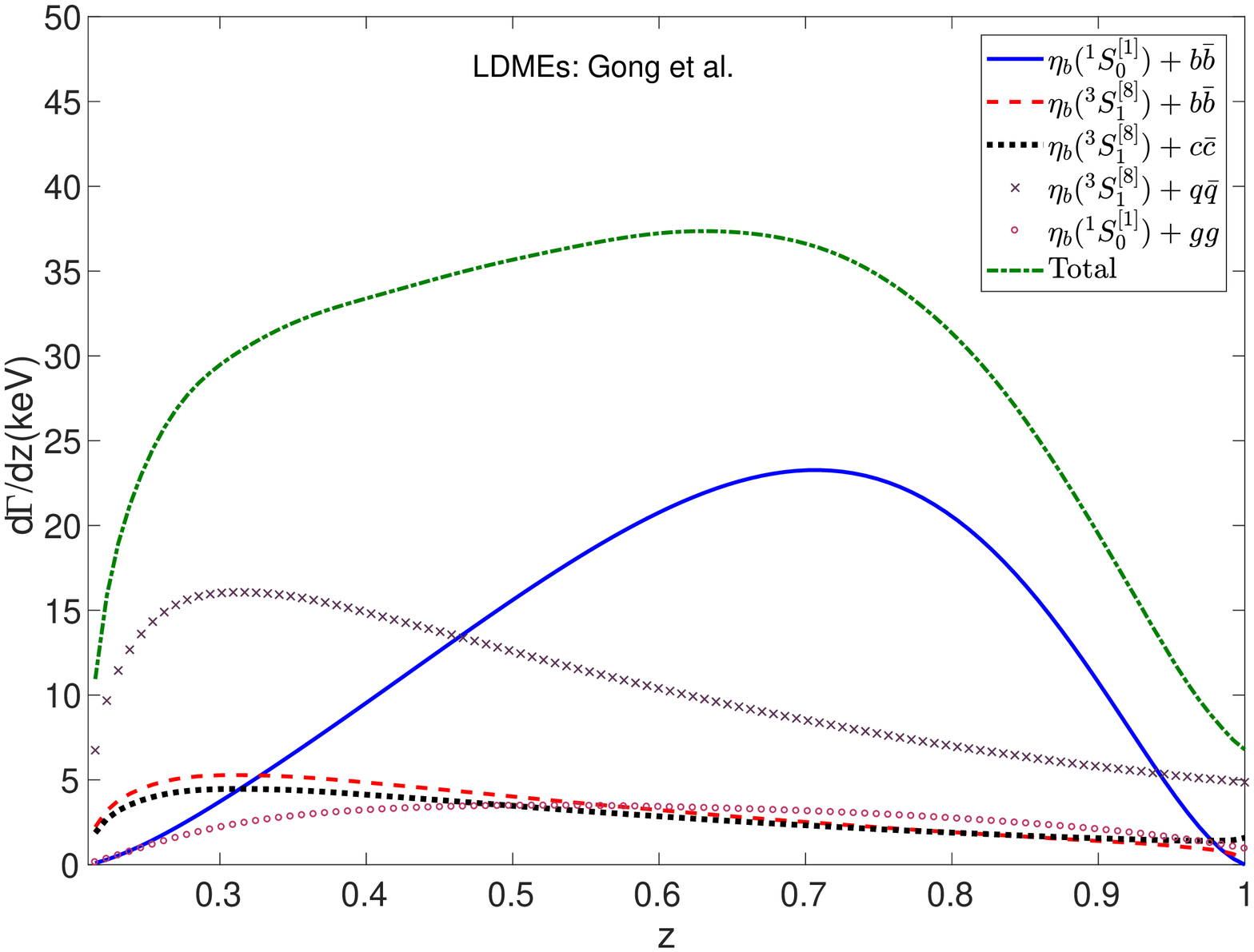}
\includegraphics[width=0.45\textwidth]{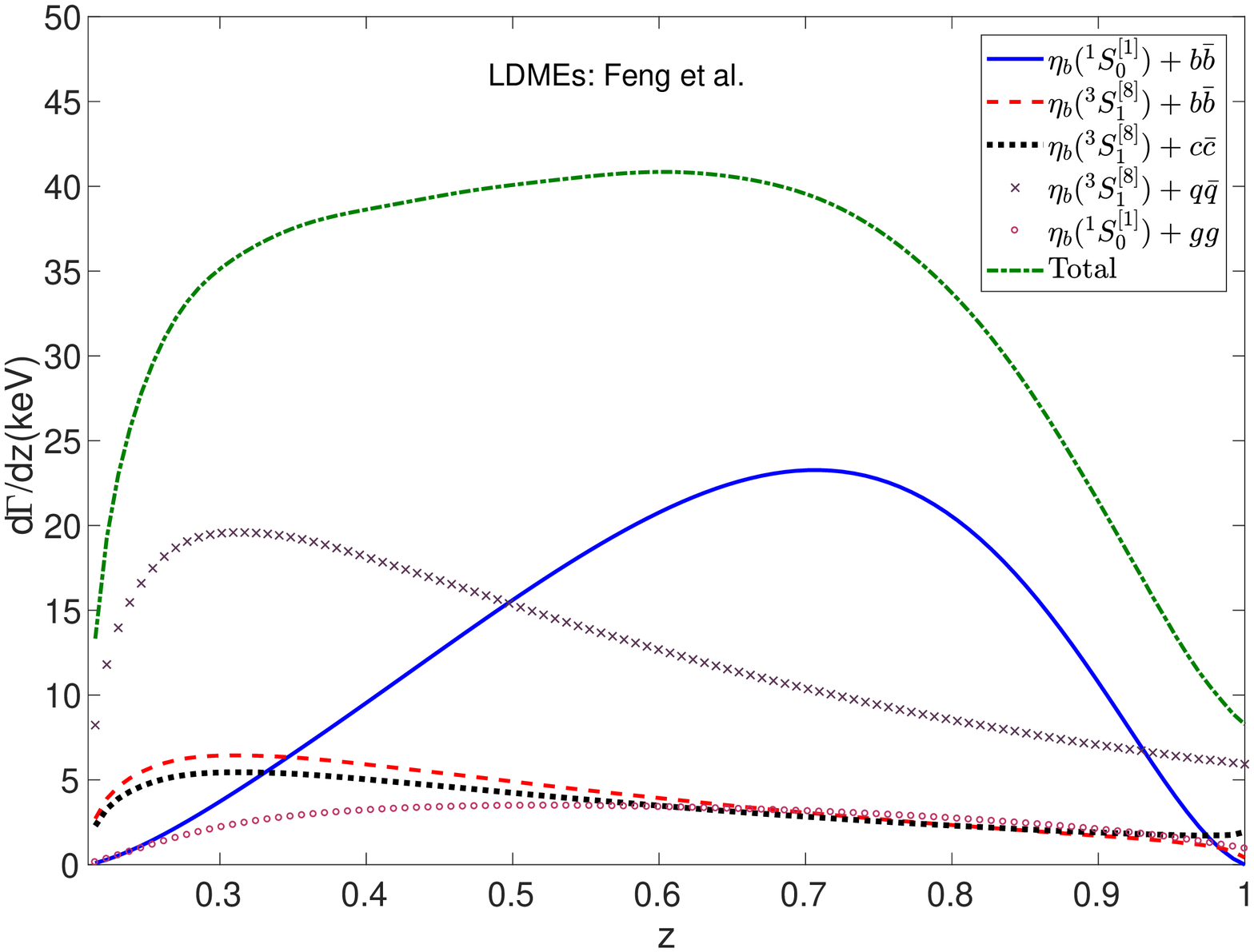}
\caption{The differential decay widths $d\Gamma/dz$ for $Z \to \eta_c+X$, where $\mu_R=2m_b$. The top one shows the distributions based on the LDMEs of Gong et al \cite{Gong:2013qka},  and the bottom one shows the distributions based the LDMEs of Feng et al \cite{Feng:2015wka}.
 } \label{gammazb}
\end{figure}

The differential decay widths for $Z \to \eta_b +X$ based on the two sets of the LDMEs are given in Fig.\ref{gammazb}. Here, the differential decay width of the channel $\eta_b(^1S_0^{[1]})+gg$ is also given. Similar to the $\eta_c$ case, the curves of the CS channels have a peak at a moderate $z$ value while the curves of the CO channels have a peak at a small $z$ value. However, since the CS contribution is comparable with the CO contribution in the $\eta_b$ case, the shape of the inclusive process $Z \to \eta_b +X$ is significantly different from that of $Z \to \eta_c +X$.

\section{Discussion and conclusion}
\label{secSum}

In the present paper, we have studied the inclusive production of $\eta_Q$ (Q=c or b) through $Z$ boson decays. The CS ($^1S_0^{[1]}$) and the CO ($^1S_0^{[8]}$, $^3S_1^{[8]}$, and $^1P_1^{[8]}$) Fock states are considered. The integrated and differential decay widths for the related channels are computed, and the results show that the decay width of $Z \to \eta_Q +X$ is dominated by the CO $^3S_1^{[8]}$ production. It means that the decay width of $Z \to \eta_Q +X$ is sensitive to the value of the LDME $\langle {\cal O}^{\eta_{Q}}(^{3}S_{1}^{[8]})\rangle$. Hence, the two processes $Z \to \eta_{c} +X$ and $Z \to \eta_{b} +X$ can be used to determine the values of $\langle {\cal O}^{\eta_{c}}(^{3}S_{1}^{[8]})\rangle$ and $\langle {\cal O}^{\eta_{b}}(^{3}S_{1}^{[8]})\rangle$. Moreover, via HQSS, the measured value of $\langle {\cal O}^{\eta_{Q}}(^{3}S_{1}^{[8]})\rangle$ from the process $Z \to \eta_Q+X$ can also give a certain constraint on the value of $\langle {\cal O}^{J/\psi(\Upsilon)}(^{1}S_{0}^{[8]})\rangle$. Note that this conclusion depends partly on the exact values of LDMEs, and it is applicable only if the LDMEs are at the same order of magnitude of the ones quoted in this paper.

The differential distributions $d\Gamma/dz$ are shown in figures. The distributions of the CS and the CO components are very different. The distributions of the CS component have a peak at a moderate $z$ value, while the distributions of the CO components have a peak at a small $z$ value. Thus, the CS LDME $\langle {\cal O}^{\eta_Q}(^{1}S_{0}^{[1]})\rangle$ and the CO LDME $\langle {\cal O}^{\eta_Q}(^{3}S_{1}^{[8]})\rangle$ can be determined more precisely through measuring the energy distribution of the process $Z \to \eta_Q +X$.

In a hadronic collider such as the LHC, the production of the heavy quarkonium $\eta_Q$ is dominated by hadronic production, so it is difficult to pick up the production events via $Z$ decays. Thus the calculations on the production via $Z$ boson decays here may be really useful as reference mainly for the production in a super $Z$ factory. The total cross section for the $\eta_Q$ production via the electron and positron annihilation at the $Z$ pole, $e^+e^- \to Z \to \eta_Q+X$~\footnote{The $\gamma$-exchange contribution is negligibly small at the $Z$-pole, which can be safely neglected.} can be derived from the decay width $\Gamma_{Z \to \eta_Q+X}$ through the formula derived in the Appendix A1 of Ref.\cite{Zheng:2017xgj}, i.e.,
\begin{eqnarray}
\sigma_{e^+e^-  \to \eta_Q+X}=\frac{e^2(1-4{\rm sin}^2\theta_W + 8{\rm sin}^4\theta_W)}{8{\rm sin}^2\theta_W {\rm cos}^2\theta_W m_{_Z} \Gamma_Z^2}\Gamma_{Z  \to \eta_Q+X}.\nonumber \\
\end{eqnarray}
Then we obtain
\begin{eqnarray}
\sigma_{e^+e^-  \to \eta_c+X}&=&45.0\,{\rm pb},  \\
\sigma_{e^+e^-  \to \eta_b+X}&=&0.608\,{\rm pb},
\end{eqnarray}
where the input values for the LDMEs have been taken as those extracted by Chao et al \cite{Chao:2012iv} and Gong et al \cite{Gong:2013qka}, which have been presented in Tables \ref{tb.LDMEs-1} and \ref{tb.LDMEs-2}. If the luminosity of a $Z$ factory can be up to $10^{35}{\rm cm}^{-2}{\rm s}^{-1}$~\cite{zfactory}, then there are about $4.5 \times 10^{7}$ $\eta_c$ and $6.1 \times 10^{5}$ $\eta_b$ to be produced per operation year. Therefore, at a high luminosity $Z$ factory with highly rejecting backgrounds, those two production processes can be studied thoroughly.

\hspace{2cm}

\noindent {\bf Acknowledgments:} This work was supported in part by the Natural Science Foundation of China under Grants No. 11625520, No. 12005028, No. 11675239, No. 11745006, No. 11821505, No. 12075301, No. 12047564, by the China Postdoctoral Science Foundation under Grant No. 2021M693743, by the Fundamental Research Funds for the Central Universities under Grant No. 2020CQJQY-Z003, and by the Chongqing Graduate Research and Innovation Foundation under Grant No. ydstd1912.

\end{document}